**Title: Braiding lateral morphotropic grain boundary in homogeneitic oxides**

**Authors:** Shengru Chen,[1,2,†] Qinghua Zhang,[1,†] Dongke Rong,[1,†] Yue Xu,[1,2] Jinfeng Zhang,[3] Fangfang Pei,[4] He Bai,[5] Yan-Xing Shang,[1,2] Shan Lin,[1,2] Qiao Jin,[1,2] Haitao Hong,[1,2] Can Wang,[1,2,6] Wensheng Yan,[4] Haizhong Guo,[7] Tao Zhu,[1,5,6] Lin Gu,[8] Yu Gong,[9] Qian Li,[4] Lingfei Wang,[3] Gang-Qin Liu,[1,2,6] Kui-juan Jin,[1,2,6,*] and Er-Jia Guo[1,2,6,*]

**Affiliations:**
[1] Beijing National Laboratory for Condensed Matter Physics and Institute of Physics, Chinese Academy of Sciences, Beijing 100190, China
[2] Department of Physics & Center of Materials Science and Optoelectronics Engineering, University of Chinese Academy of Sciences, Beijing 100049, China
[3] Hefei National Laboratory for Physical Science at the Microscale, University of Science and Technology of China, Hefei 230026, China
[4] National Synchrotron Radiation Laboratory, University of Science and Technology of China, Hefei 230029, China
[5] Spallation Neutron Source Science Center, Dongguan 523803, China
[6] Songshan Lake Materials Laboratory, Dongguan, Guangdong 523808, China
[7] Key Laboratory of Material Physics & School of Physics and Microelectronics, Zhengzhou University, Zhengzhou 450001, China
[8] National Center for Electron Microscopy in Beijing and School of Materials Science and Engineering, Tsinghua University, Beijing 100084, China
[9] Department of Physics and Astronomy, College of Charleston, 58 Coming Street, Charleston, SC 29424, USA

[†]These authors contribute equally to the manuscript.
*Corresponding author. Emails: kjjin@iphy.ac.cn and ejguo@iphy.ac.cn

**Abstract:** Interfaces formed by correlated oxides offer a critical avenue for discovering emergent phenomena and quantum states. However, the fabrication of oxide interfaces with variable crystallographic orientations and strain states integrated along a film plane is extremely challenge by conventional layer-by-layer stacking or self-assembling. Here, we report the creation of morphotropic grain boundaries (GBs) in laterally interconnected cobaltite homostructures. Single-crystalline substrates and suspended ultrathin freestanding membranes provide independent templates for coherent epitaxy and constraint on the growth orientation, resulting in seamless and atomically sharp GBs. Electronic states and magnetic behavior in hybrid structures are laterally modulated and isolated by GBs, enabling artificially engineered functionalities in the planar matrix. Our work offers a simple and scalable method for fabricating unprecedented innovative interfaces through controlled synthesis routes as well as provides a platform for exploring potential applications in neuromorphics, solid state batteries, and catalysis.

**One-Sentence Summary:** Lateral-engineered morphotropic grain boundaries are achieved in homogeneitic cobaltite films, creating an avenue for innovative functionalities.

**Main Text:**

Advanced thin-film deposition techniques enable scientists to fabricate high-quality oxide interfaces exhibiting emergent phenomena not found in bulk constituents (*1-3*). Interfaces naturally break spatial inversion symmetry, resulting in the reconstruction of electronic degrees of freedom at the nanoscale (*4, 5*). These features typically improve the effects of electron correlations such as the reduction of bandwidths and dispersion, changes in crystal fields, and shifts of Madelung energies (*6*), significantly impacting macroscopic physical properties. Therefore, interfaces have several intriguing and potentially useful effects, including high-temperature ferromagnetism, high-$T_C$ superconductivity, electronic phase transitions, and multiferroicity (*7-10*). Although a broad range of oxide interfaces have been synthesized and maturely investigated over decades, the conventional epitaxy focuses primarily on functional multilayers and superlattices stacked vertically layer-by-layer (*11-13*). High-quality epitaxial growth occurs only between dissimilar materials with similar crystal structures and close lattice constants. Beyond certain thresholds, an increasing number of defects and dislocations will form at interfaces, severely degrading their functionalities. In such cases, underlying substrates determine the crystallographic orientation, symmetry, and misfit strain of epitaxial layers grown on the surface. Another type of interface is aligned vertically in self-assembled arrayed nanostructures (*14*). Intrinsic similarities in the crystal chemistry, *e. g.*, oxygen coordination and electronegativity, allow the tantalizing possibility of epitaxy in three dimensions. Nanoscale clusters of ferro-/ferri-magnetic materials embedded in a ferroelectric matrix exhibit strong mechanical coupling between two ferroic order parameters, piquing a growing interest in developing multiferroic devices based on composite heterostructures (*15*). Success in these endeavors unquestionably promote the comprehensive understanding of correlated electron materials as well as their incorporation into practical applications.

Developing feasible strategies for fabricating innovative oxide interfaces provides many opportunities in both fundamental research and evolutionary devices. Lateral homostructures exhibit in-plane spatial modulations of the structural and physical properties of an identical single crystal in contrast to well-defined longitudinal heterostructures. One famous example is the formation of a morphotropic phase boundary in bismuth ferrite ($BiFeO_3$) films through epitaxial constrain (*16*). Both tetragonal- and rhombohedral-like phases coexist within a single film and can be reversibly switched by an electric field and mechanical force (*16, 17*). The boundary between two distinct phases significantly modulates surface morphology and exhibits enhanced domain wall conductance in topologically confined nanoislands, implying great potential in non-volatile memory and actuator applications (*18*). In-plane spatially patterned electronic textures can also be achieved in ferromagnetic manganite and cobaltite single films by combining surface miscut steps and domain patterns via striped strain modulation. (*19, 20*). Furthermore, modification to the lattice symmetry in the film plane has been established in oxide films through low energy helium implantation post growth (*21, 22*). This non-destructive strain doping approach is readily applied locally down to the micrometer scale to the same sample, effectively altering the crystal's total anisotropy energy and magnetic properties. From the traditional thin-film epitaxy perspective, the growth of lateral phase boundaries in homostructures composed of various crystallographic orientations and in-plane strain states remains a significant challenge. To remove the constrain regularity from single crystal substrates, a strategy that bypasses these restrictions in thin-film epitaxy is highly required. We noticed that a recent work by Wu *et al*. report the growth of twisted multiferroic oxide lateral homostructures using freestanding membranes (*23*). The functional oxide layers can be rotated and stacked with designed twisting angles, adding another degree of freedom to design thin film properties. However, the fabrication method involves the hydrochloric acid as etching materials. The unavoidable chemical residuals and relatively large surface roughness of transferred membranes result in a great challenge in fabricating atomic-thin spatial gap between membrane and substrate. These factors inevitably influence the flatness, crystallinity, and adhesiveness of subsequent grown layers, as well as the sharpness of morphotropic grain boundaries (GBs). Recent progresses in the synthesis of atomically thin membranes using water-soluble sacrificial layers allow us to develop ultrathin freestanding membranes with ultra-sharp interfaces without using highly corrosive acid.

Here, we applied chemical free and environmental-friendly strategy in which only the deionized water is required to synthesize ~ 8-unit-cell thick freestanding single crystal membranes. These high-quality membranes attached firmly onto the substrates maintain a minimum spatial gap down to ~ 1 nm without forming chemical bonds, allowing these membranes to maintain their original structural geometries. Ferromagnetic oxide films epitaxially grown on both membranes and substrates strictly follow the respective regularity independently. Atomically sharp interfaces are formed at morphotropic GBs between homogeneitic layers with different orientations and strain states. The atomic-thin GBs naturally isolate the distinct regions with strikingly different electronic states and magnetic behavior in hybrid structures, enabling artificially engineered functionalities in the planar matrix. The ability to fine-tune lateral structure-driven functional properties demonstrates a practical path toward the rational design of magnetoelectric and spintronic logic devices, as well as microscopic surface catalysis.

**Results**

The water-soluble sacrificial strontium aluminate ($Sr_3Al_2O_6$) layer (~ 30 nm) and a subsequent ultrathin $SrTiO_3$ (STO) layer (~ 3 nm) were grown on a (001)-oriented $(LaAlO_3)_{0.3}$-$(Sr_2AlTaO_6)_{0.7}$ (LSAT) single-crystalline substrate using the pulsed laser deposition technique. Then, the STO layer was attached to a thermal-release tape and delaminated from LSAT substrates after dissolving the sacrificial layer in de-ionized water (*24, 25, 26*). The released single-crystalline STO membranes were transferred onto the target substrates (see Materials and Methods and fig. S1). The STO membranes maintain their shape and crystalline quality in millimeter size. We did not cover the entire surface of the substrates with the STO membranes to allow subsequent thin layers to be epitaxially grown on both regions simultaneously. The area ratio between STO membranes and substrates was 1:3. We chose $La_{0.8}Sr_{0.2}CoO_3$ (LSCO) as a model system because its specific Sr composition is close to the phase boundary, where small structural perturbations significantly change its physical properties (*27*). LSCO layers with a thickness of ~ 30 nm were grown on the modified substrates. A representative hybrid structure composed of (001)- and (110)-oriented LSCO layers on (110)-STO substrates is displayed in Fig. 1a. X-ray diffraction (XRD) measurements confirm the epitaxial growth of LSCO layers with both (00*l*) and (*ll*0) reflections (Fig. 1b and fig. S2). The crystalline structure of the (001)-STO membranes is preserved, and these membranes act as a template for subsequent LSCO layer growth. Conversely, the ultrathin STO membranes render them responsive to compressive strain from the top LSCO layers. The out-of-plane lattice constant of the (001)-LSCO layer (~ 3.81 Å) grown on STO membranes is larger than that of a coherently strained (001)-LSCO single film (~ 3.77 Å), implying that the lattice structure of the LSCO layer is slightly relaxed. The atomic-scale structural coherency of LSCO hybrid structures grown on (110)-oriented STO substrates was examined by cross-sectional high-angle annular dark-field (HAADF) imaging via scanning transmission electron microscopy (STEM) (Fig. 1c and fig. S3). The ultrathin STO membranes with a thickness of ~ 8 unit cells are continuous, atomically smooth, and free from chemically intermixing or defect formation after release, transfer, and post-synthesis (fig. S4). The extremely narrow gap (~ 1 nm on average) between the STO membranes and substrates is naturally formed (fig. S5). Most STO membranes are chemically unbonded with substrates because of the crystallographic orientation mismatch and polar interface. Atomic-resolution STEM images of an LSCO hybrid structure in different regions (Fig. 1d and fig. S6) show that LSCO layers strictly follow the orientations of STO membranes and substrates. The schematics of the atomic structure shown in Fig. 1d manifest the epitaxial relationship across the distinct LSCO/STO interfaces. Fig. 1e shows a HAADF-STEM image of the GB between homogeneitic LSCO layers with different orientations. Notably, the GB has an atomically sharp interface with a highly periodic configuration of atoms due to energy minimization to misfit strain and different crystallographic facets. Fig. 1Fshows the averaged atomic intensities in the HAADF-STEM image at different interface regions (Fig. 1e). This qualitative analysis enables the precise realization of out-of-plane lattice constants of each layer. We also performed a geometric phase analysis (GPA) of HAADF-STEM images across the LSCO/STO interfaces (fig. S7). Similar to the coherently epitaxial growth of (110)-LSCO layers on STO substrates, the (001)-LSCO layers have an identical in-plane lattice constant but a smaller out-of-plane lattice constant than those of STO membranes. The top LSCO layer compresses the STO membranes along the in-plane direction, increasing the out-of-plane lattice constant, compared with the STO bulk. The STEM results confirm the structural relaxation of (001)-LSCO layers on STO membranes, which is consistent with XRD measurements. Furthermore, following the same protocol, we successfully constructed a similar hybrid structure consisting of (001)- and (111)-oriented LSCO layers on (111)-STO substrates (fig. S8), confirming that our methodology for fabricating homostructures with lateral GBs is universal and highly reproducible.

In addition to the detailed analysis of structural evolution and GBs, we investigated the transport and magnetic properties of LSCO hybrid structures using (110)-STO substrates. For a (001)-LSCO/STO freestanding (FS) membrane attached to PDMS support, the in-plane saturation magnetization ($M_S$) is ~ 200 emu/cm$^3$ and the coercive field ($H_C$) is ~ 5 kOe at 10 K (Fig. 2a). $M_S$ reduces to 85 emu/cm$^3$, and $H_C$ increases to ~ 8 kOe when an LSCO single film is grown on (110)-STO substrates (Fig. 2b). A strong reduction in the ordered magnetic moment with increasing tensile strain is consistent with earlier work (28). The decrease in $M_S$ is irrelevant to the change in Co covalence and spin state because the tensile strain increases the magnetic moment in undoped cobaltite films (29). We observed a two-step switching in the magnetic hysteresis loop of an LSCO hybrid structure grown on (110)-STO substrates (Fig. 2c), implying that the (001)-STO and (110)-STO layers respond differently to the applied fields. Fig. 2d shows temperature-dependent resistivity ($\rho$) in a (001)-FS-LSCO, a (110)-LSCO single layer, and an LSCO hybrid structure. (001)-FS-LSCO is highly insulating, with its $\rho$ exceeding the measurement limit at room temperature ($\rho_{300K} > 100$ Ω·cm). The $\rho$ of an LSCO hybrid structure is smaller than that of a (110)-LSCO single film, exhibiting a typical resistance property of the parallel-connected circuit. The temperature-dependent magnetization ($M$) at an in-plane field of 1 kOe exhibits nearly identical $T_C$ ~ 195 K for all three samples (Fig. 2e). The $M$ of an LSCO hybrid structure is larger than that of a (110)-LSCO single layer but smaller than that of a (001)-FS-LSCO. Figs. 2f and 2g show the field-dependent magnetoresistance [$(\rho_H-\rho_0)/\rho_0$] and $M$ of an LSCO hybrid structure at various temperatures. $(\rho_H-\rho_0)/\rho_0$ and the magnetic field at peak positions (corresponding to $H_C$) reduces progressively with increasing temperature, revealing a typical ferromagnetic characteristic. The double hysteresis loop in LSCO hybrid structures disappears when $T$ exceeds 100 K. This fact suggests that, at high temperatures, the magnetic contribution from (110)-STO layers is negligible, and (001)-FS-LSCO layers dominate the total magnetization of an LSCO hybrid structure. In addition, similar field-dependent magnetization was observed in LSCO hybrid structures on (111)-STO substrates (fig. S9). These findings show that LSCO layers with different orientations embedded in a single hybrid structure have unique physical properties and regions separated by lateral GB responses that are independent of external fields.

To verify that our method can be universally used to create GBs between oxide layers with different strain states, we grew LSCO hybrid structures on STO-membrane-covered LaAlO$_3$ (LAO) and KTaO$_3$ (KTO) substrates with a (001) orientation, respectively. The in-plane strain of LSCO layers varies from compressive (~ −1.25% on LAO) to tensile (~1.7% on STO and −4% on KTO) depending on substrates (Fig. 3d). We firstly verified that a compressively strained LSCO single film on LAO undergoes an insulator-to-metal transition at ~ 175 K, whereas the tensile-strained LSCO films on STO and KTO exhibit insulating behavior at all temperatures (fig. S10). $M_S$ at 10 K under a magnetic field of 7 T decreases significantly with increasing in-plane lattice parameters. Through polarized neutron reflectometry, we also confirm that the magnetization of an LSCO single film is uniform across the entire sample and only a small variation (less than 5%) exists at the surface and interface (fig. S11) due to the different boundary conditions (*30*). Therefore, LSCO is an excellent candidate for manipulating strain-driven electronic and magnetic properties across GBs. Fig. 3a shows a high-resolution HAADF-STEM image across interfaces between LSCO layers, FS-STO membranes, and LAO substrates. The [010] orientation of FS-STO membranes is twisted ~ 16° with respect to the [010]$_{pc}$ orientation of LAO substrates (fig. S12), which is confirmed by the atomic-resolution STEM image from top LSCO/STO layers, but well-aligned atomic planes at LAO substrates. From GPA and line profiles across heterointerfaces, the lattice constants of LSCO layers can be accurately obtained (Figs. 3b and 3c, fig. S13). Both LSCO layers are coherently grown on both LAO substrates and STO membranes. The LSCO layers undergo compressive strain caused by LAO, whereas they are slightly tensile-strained by STO membranes. The STO membranes are conversely in-plane compressed by LSCO top layers, resulting in a slightly higher out-of-plane lattice constant (~ 3.92 Å) than its bulk value.

Distinct strain states of LSCO layers in different regions would induce significant lateral modulation in the orbital occupancy of valence electrons. We performed elemental-specific X-ray absorption spectroscopy (XAS) at Co $L$-edges on an LSCO hybrid structure. XAS measurements were conducted at 77 K under a zero magnetic field, ensuring that XAS reflects the genuine electronic occupancy of Co $d$ orbitals. Fig. 3e shows the experiment configuration for XAS measurements at the edge (LSCO/LAO) and center (LSCO/STO) regions of an LSCO hybrid structure. By detecting the absorption of linearly polarized X-rays with different incident angles ($I_{0°}$ and $I_{60°}$), we could probe the respective energies and unoccupied states, *i. e.,* holes, in the Co $d_{x^2-y^2}$ and $d_{3z^2-r^2}$ orbitals (fig. S14). X-ray linear dichroism (XLD) spectra were quantified by calculating the difference between $I_{0°}$ and $I_{60°}$, directly reflecting the orbital polarization in different regions of an LSCO hybrid structure. As shown in Fig. 3f, XLD spectra from edge and center regions of an LSCO hybrid structure on LAO exhibit an obvious reversed sign. Compressive strain-induced orbital splitting results in low energy and highly occupied $d_{3z^2-r^2}$ orbitals, whereas tensile strain has the opposite effect. Furthermore, the XLD of an LSCO hybrid structure on KTO confirms that electrons preferentially occupy the $d_{x^2-y^2}$ orbitals when LSCO layers are tensile-strained by STO membranes or KTO substrates (Fig. 3g). The XLD results are consistent with those obtained for LSCO single films grown on the respective substrates (fig. S15), showing that strain-driven orbital occupancy switching is robust (*31*). These observations provide strong evidence that lateral strain modulation induces significant in-plane anisotropy in the electronic configuration, affecting both band splitting and orbital polarization in the Co *d* bands.

The effects of lateral strain modulation and GBs on the electric and magnetic properties of LSCO hybrid structures were further investigated by DC transport measurements. Magnetoresistance and Hall conductivity were measured at the edge (Fig. 4a) and center (Fig. 4b) regions of an LSCO hybrid structure independently (figs. S16 and S17). Temperature-dependent *ρ* of LSCO hybrid structures exhibit a semimetallic behavior with a local maximum at ~ 170 K, corresponding to the magnetic phase transition temperature. Comparing this behavior with an LSCO single film on LAO, we observe quantitatively similar temperature-dependent resistivity that is roughly an order of magnitude larger than LSCO hybrid structures. This is due to the highly insulating state in tensile-strained LSCO layers on STO membranes, leading to finite electrical conductivity. The Hall measurements of LSCO hybrid structures led to an unprecedented observation. Above $T_C$, the Hall response is approximately linear with a positive slope, indicating *p*-type carriers. At low temperatures, a nonlinear anomalous Hall effect (AHE) appears, opening a large hysteresis loop with a shape and coercive field practically identical to *M-H* sweeps. When *T* is below 30 K, AHE disappears. Figs. 4c and 4d show direct comparisons of anomalous Hall resistance ($\rho_{xy}$-$R_0H$) in different regions as a function of magnetic field and temperature, respectively. At 50 K, $\rho_{xy}$-$R_0H$ (edge region) has a square-like loop, whereas $\rho_{xy}$-$R_0H$ (center region) has negligible values. $\rho_{xy}$-$R_0H$ reaches a maximum (minimum) value at 100 K when measured at edge (center) regions; however, the numerical sign of $\rho_{xy}$-$R_0H$ is the opposite. In addition, the magnetoresistance of LSCO hybrid structures at both edge and center regions has negligible values at high temperatures. A negative response starts to appear near and below the $T_C$ and increases systematically in magnitude with decreasing temperature. When *T* is below 100 K, the butterfly hysteresis loop appears in the small field that is coupled with $H_C$. Figs. 4e and 4f plot field- and temperature-dependent magnetoresistance recorded in different regions, respectively. They show a distinct response to the applied field below $T_C$, indicating that the onset of magnetization in LSCO hybrid structures is laterally different.

The direct evidence of the magnetic contrast across the GB is provided by nanodiamond (ND) nitrogen vacancy (NV) magnetometry measurements at the edge and center regions of an LSCO hybrid structure on LAO substrates. NV magnetometry was performed in zero magnetic fields and at fixed temperatures ranging from 50 K to 250 K. By recording optically detected magnetic resonance (ODMR) spectra of NV centers, we could calculate the projection of the magnetic stray field that originated from LSCO layers along the NV axis because $m_s = \pm 1$ sublevels undergo energy splitting ($2\gamma B$) in the presence of weak magnetic perturbation (*B*) (Fig. 4g). Fig. 4h and fig. S18 depicts ODMR spectra of a single ND dispersed on the surface of edge and center regions at various temperatures. At 50 K, the splitting between two resonant peaks is significant, indicating the existence of measurable remnant magnetization in LSCO layers. The splitting becomes narrow and reaches a saturated minimum value as temperature increases. We calculated $2\gamma B$ by subtracting two resonant peak frequencies after Gaussian fitting to ODMR spectra. Fig. 4i shows temperature-dependent $2\gamma B$ measured at the edge and center regions of LSCO hybrid structures. The magnitude of $2\gamma B$ depends on multiple factors, including the number of NVs in a single ND, proximity distance, and magnetic homogeneity. Thus, the quantitative comparison of $2\gamma B$ obtained from different regions is challenging. However, the magnetic phase transition temperatures of LSCO layers at edge and center regions are significantly different. LSCO (edge) layers on LAO substrates have a lower $T_C$ than LSCO (center) layers on STO membranes. These measurements were repeated randomly using different NDs in each region. The trend in magnetization obtained from NV magnetometry correlates with $M$-$T$ curves from macroscopic magnetometry results on LSCO single films and LSCO/STO membranes (Fig. 2e, figs. S9 and S10), corroborating the lateral magnetic anisotropy of LSCO hybrid structures with different strain states isolated by GBs.

**Discussion and conclusions**

So far, we have demonstrated the creating of lateral GBs in ferromagnetic LSCO hybrid structures with different crystallographic orientations and strain states. By designing and artificially engineering GBs in a controlled way, this work may also open all kinds of interesting fundamental and applied research out of quantum materials. The results are highly reproducible, and most importantly, the growth sequence can be reversed (fig. S19). For instance, we fabricated (110)- and (111)-oriented STO membranes and attached them to (001)-STO substrates. Hence, LSCO hybrid structures were formed on (001)-STO substrates, which exhibit similar magnetic contrast laterally. In principle, the freestanding membranes (not limited to STO used in this study) can be transferred onto arbitrary substrates, such as metals, semiconductors, as well as amorphous glass, which is crucial for fabricating complex laterally hybrid structures with emergent physical properties (fig. S19). This work can be applied to a broad range of applications. First, the size of membranes can be reduced to a nanometer scale using a sacrificed nanoporous anodic alumina template or through UV lithography post-synthesis (*32*). The arrayed nanodots enable on-demand manipulation of topological nanoripples, spin-textured domains, or multiferroic vortices. Second, the creation of GBs composed of unique periodic arrangements of structural units in any single-crystalline oxide is achievable and accurately adjustable by function-driven design. The type, density and location of GBs is highly controllable. Tunability via GB engineering with broken inversion symmetry may open vistas in nanoelectronics and nanoelectromechanical systems (*33*). Last but not the least, the proposed technique enables the stacking of correlated oxide membranes with precisely controlled twisted angles with respect to subsequent-grown layers, mimicking twistronics in two-dimensional materials (*34*). This method advances our ability to minutely engineer functional properties by lateral homoepitaxy of strongly correlated materials that drive many intriguing behaviors, such as superconductivity, multiferroicity, and magnetic textures, across multiple length scales, and is readily applied to the imperative fields of neuromorphics, solid state batteries, catalysis, *etc*.

**Materials and Methods**

**Synthesis of LSCO hybrid structures**

The water-soluble $Sr_3Al_2O_6$ (SAO) and $SrTiO_3$ (STO) layers were grown subsequently on (001)-oriented $(LaAlO_3)_{0.3}$-$(Sr_2AlTaO_6)_{0.7}$ (LSAT) substrates (Hefei Kejing Mater. Tech. Co. Ltd) by pulsed laser deposition (PLD) technique. A focused XeCl excimer laser with duration of ~ 25 ns, fixed wavelength of 308 nm, and energy density of ~ 1.5 J/cm$^2$ was used as the ablation source. The bilayers were deposited at the substrate temperature of 800 °C and oxygen partial pressure of 50 mTorr. The thicknesses of SAO and STO layers were ~ 30 nm and ~ 3 nm (approximately ~ 8 unit cells, u. c.), respectively. On the completion of epitaxy, the STO/SAO bilayers were cooled down to room temperature at the growth pressure. A thermal-release tape (or a PDMS) was pressed firmly on the as-grown sample and then immersed into de-ionized water at room temperature (fig. S1). After the water-soluble SAO layer was fully dissolved, the ultrathin STO membrane was adhered on the thermal-release tape (or a PDMS). Afterwards, the tape-supported STO membranes were transferred on the target substrates, for instance the (110)- and (111)-oriented STO, LaAlO$_3$ (LAO), KTaO$_3$ (KTO), and SiO$_2$/Si substrates, followed by peeling off the thermal-release tape by heating at ~ 90 $^o$C for 10 minutes. After that, the ultrathin STO membranes remained on the target substrates after detaching the supports. Subsequently, the La$_{0.8}$Sr$_{0.2}$CoO$_3$ (LSCO) thin films with a thickness of ~ 30 nm were fabricated on the prepared substrates by PLD. The thickness of LSCO films was carefully selected by keeping them coherently grown on different target substrates. The LSCO films exhibit distinct physical properties depending on the crystallographic orientation and misfit strain. During the growth of LSCO layers, the substrate temperature was kept at 750 $^o$C and oxygen partial pressure was maintained at 200 mTorr. After the deposition, the samples were cooled down under the oxygen pressure of 100 Torr. The post-oxygen annealing process prevents the non-stoichiometry in the LSCO films that would possibly affect their intrinsic physical properties.

**Structural characterizations**

Crystallographic analysis, that is, X-ray diffraction 2$\theta$-$\omega$, X-ray reflectivity (XRR), and reciprocal space mapping (RSM), were carried out using a Panalytical X'Pert3 MRD diffractometer with Cu $K\alpha_1$ radiation equipped with a 3D pixel detector. The thicknesses of layers were obtained by fitting XRR curves using GenX software (*34*). The growth rate of each layer is calculated and controlled precisely by counting the number of laser pulses. Cross-sectional TEM specimens of LSCO hybrid structures with different crystallographic and strain states were prepared using Ga$^+$ ion milling after the mechanical thinning. The HAADF and ABF imaging were performed in the scanning mode using JEM ARM 200CF microscopy at the Institute of Physics (IOP), Chinese Academy of Sciences (CAS). For some cases, the [100] zone axis of STO(001) membranes is not well aligned with [$\bar{1}$10] or [11$\bar{2}$] zone axis of STO(011) or STO (111) substrates, respectively. The LSCO hybrid specimens need to rotate by a few degrees along the in-plane direction in order to obtain the atomic-precision images from different regions. The atomic positions of cations were determined by fitting the intensity peaks with Gaussian function. The obtained values were used to calculate the lattice parameters of each layer. The error bars were extracted by calculating the standard deviation value. The elemental specific electron energy loss spectra (EELS) mappings were performed by integrating the signals from a selected region after subtracting the exponent background using power law. All data were analyzed using Gatan Micrograph software.

**Electrical transport and magnetization measurements**

The macroscopic magnetizations of LSCO hybrid structures were measured by a 7 T−MPMS magnetometer. All measurements were performed by applying magnetic fields up to +/− 7 T. The *M-T* curves were recorded during the sample warm-up process after field-cooled at 1 kOe. The *M-H* hysteresis loops up to +/− 7 T were recorded at different temperatures. The *M-H* loops were obtained by subtracting the diamagnetic signals from the STO membranes and substrates. The magnetization of LSCO hybrid structures were normalized to the thickness of LSCO layers. The electrical transport measurements were conducted by a 7 T−PPMS. The resistivities and Hall conductivities were measured using the standard van der Paw method.

**Spectroscopic characterizations**

Elemental specific XAS measurements were performed on the LSCO hybrid structures grown on LAO and KTO substrates at the MCD beamline of National Synchrotron Radiation Laboratory (NSRL) in Hefei, China. All spectra at the Co *L*-edges were collected at 77 K in total electron yield (TEY) mode. The XLD measurements were performed by changing the incident angle of the linearly polarized X-ray beam. The X-ray scattering plane was rotated by 0° and 60° with respect to the incoming photons. When the X-ray beam is perpendicular to the surface plane (0º), the XAS signal directly reflects the $d_{x^2-y^2}$ orbital occupancy. While the angle between the X-ray beam and surface plane is 60°, the XAS signal contains orbital information from both $d_{x^2-y^2}$ and $d_{3z^2-r^2}$ orbitals. For simplifying the discussions, the XLD signals of LSCO hybrid structures were calculated by $I_{0°} - I_{60°}$. The XLD signal directly reflects the orbital polarization of a sample under different strain states (*35*).

**Diamond NV-based magnetometry**

The diamond nitrogen vacancy (NV)-based magnetometry measurements were performed at zero magnetic field using a home-built optically detected magnetic resonance (ODMR) system (*36*). Nanodiamonds (NDs) with a nominal diameter of ~ 100 nm and typically ~500 NV centers per crystal were used (Adámas Nanotechnologies) as spin sensors. We dispersed the NDs on the surface of LSCO hybrid structures randomly with a low density so that the individual NDs could be addressed and probed with the confocal microscope. The ODMR spectra were taken by sweeping the microwave frequency through resonance and recording the photon excitations from the NDs. The ground states of NV center are spin-1 and there is only one resonant dip at 2870 MHz at zero magnetic field ($m_S = \pm 1$ states are degenerated) and room temperature. If a stray field is generated by the LSCO hybrids, the ODMR spectrum of an NV center splits into two dips due to the Zeeman effect. The amplitude of spectral splitting is proportion to the strength of the stray field. Thus, we extracted the local field strength from the width of single-peak Lorentz fitting. Please note that there are hundreds of NV centers in a single ND and the stray field may have a gradient at the 100-nm scale, thus the ODMR spectra of NDs from our LSCO hybrids are broadened as compared to that of a single NV center. The samples were pasted on a cooling stage attached to a closure He-recycling refrigerator. The temperature of the samples was controlled between 10 to 300 K and monitored using conventional thermometer attached on the sample stage. The cooling stage can be moved precisely at the micrometer scale, so that the ODMR spectra and their temperature dependencies from different regions of a LSCO hybrid structure can be obtained separately.

**Nanoscale magnetization profiling using PNR**

The PNR experiment on a LSCO single film grown on the LAO substrate was performed at the Multipurpose Reflectometer (MR) of Chinese Spallation Neutron Source, Dongguan, China. The sample was measured at 10 K under an in-plane magnetic field of 1 T. PNR measurements were conducted in the specular reflection geometry with wave vector transfer ($q$) perpendicular to the surface plane. $q$ is calculated by $4\pi \sin(\alpha_i)/\lambda$, where $\alpha_i$ is the neutron incident angle and $\lambda$ is the wavelength of neutron beam. The neutron reflectivities from spin-up ($R^+$) and spin down ($R^-$) neutrons were recorded separately. The spin asymmetry (SA) was calculated using $(R^+ - R^-)/(R^+ + R^-)$. By fitting the PNR data, we obtain the magnetization and nuclear scattering length density profiles of LSCO layers simultaneously. The standard deviations of the magnetization values form the uncertainties of the *M*s.

**Acknowledgments:** We thank Xiahan Sang, T. Zac Ward, Michael R. Fitzsimmons, Sujit Das, Guoqiang Yu, Yu Ye, Qian Li, Hangwen Guo, and Qiyang Lu for valuable discussions, as well as Tengyu Guo and Baoshan Cui for the extensive assistance in the magnetization measurements at SLAB. **Funding**: This work was supported by the National Key Basic Research Program of China (Grant Nos. 2020YFA0309100 and 2019YFA0308500), the National Natural Science Foundation of China (Grant Nos. 11974390, 11721404, 12174364, 11874412, and 12174437), the Beijing Nova Program of Science and Technology (Grant No. Z191100001119112), the Beijing Natural Science Foundation (Grant No. 2202060), the Guangdong-Hong Kong-Macao Joint Laboratory for Neutron Scattering Science and Technology, the Strategic Priority Research Program (B) of the Chinese Academy of Sciences (Grant No. XDB33030200), Excellence Program of Hefei Science Center CAS (No. 2021HSC-UE003), and the Fundamental Research Funds for the Central Universities (No. wk2310000104). The XAS and XLD experiments were conducted at the National Synchrotron Radiation Laboratory (NSRL) in Hefei, China via a user proposal. **Author contributions:** The LSCO samples were grown and processed by S.R.C. under the guidance of E.J.G., and the SAO samples were fabricated by J.F.Z. under the guidance of L.F.W.; the transfer of freestanding oxide membranes was conducted by D.K.R.; XAS and XLD measurements were performed by F.F.Y., Q.L., and W.S.Y.; the NV magnetometry was performed by Y.X., Y.X.S., and G.Q.L.; TEM lamellas were fabricated with FIB milling and TEM experiments were performed by Q.H.Z. and L.G.; PNR measurements were performed by H.B. and T.Z.; S.R.C., S.L., Q.J., and H.T.H. worked on the structural and magnetic measurements. C.W. and H.G. participated the discussions and K.J.J. provided important suggestions during the manuscript preparation. E.J.G. initiated the research and supervised the work. S.R.C. and E.J.G. wrote the manuscript with inputs from all authors. **Competing interests:** The authors declare that they have no competing financial interests. **Data and materials availability:** The data that support the findings of this study are available on the proper request from the first author (S.R.C.) and the corresponding authors (E.J.G.). **Additional information**: Supplementary information is available in the online version of the paper. Reprints and permissions information is available online. Certain commercial equipment is identified in this paper to foster understanding.

**Supplementary Materials**

Figs. S1 to S19

References (*34–38*)

**Figures and figure captions**

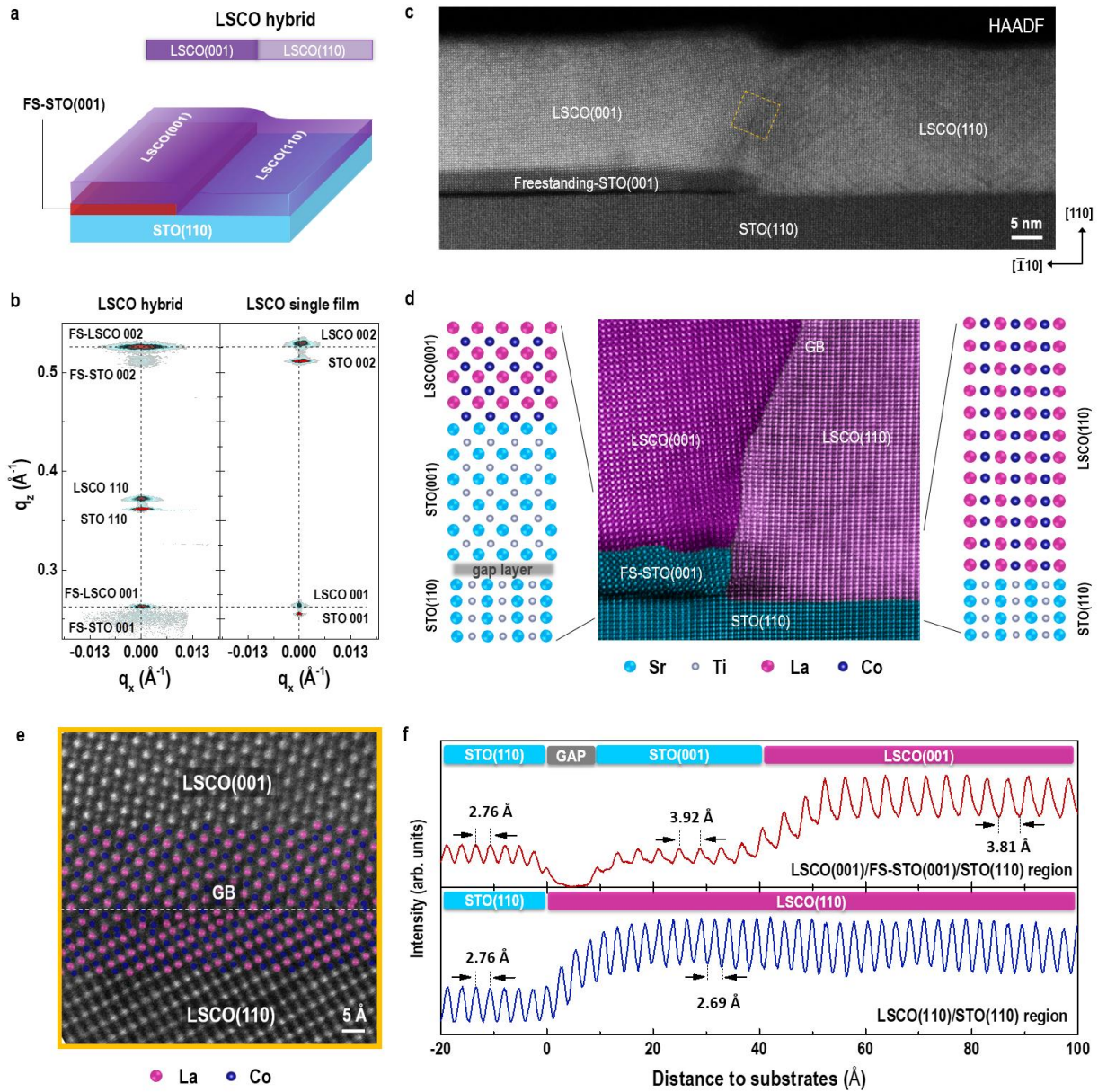

**Fig. 1. Structural characterization of a LSCO hybrid structure with morphotropic grain boundaries.** (**a**) Schematic of a lateral homogeneitic LSCO hybrid structure. The LSCO layers grown on (110)-oriented STO substrates maintain the substrate's orientation, whereas the LSCO layers follow the orientation of (001)-oriented freestanding (FS)-STO membranes underneath. (**b**) Reciprocal space map (RSM) of a LSCO hybrid structure and a (001)-oriented LSCO single layer around the substrates' (110) reflection. Both (00$l$) reflections from FS-STO and LSCO layers are observed besides the typical (110) reflections from the LSCO film and substrates. (**c**) Cross-sectional high-angle annular dark-field (HAADF) STEM images of a LSCO hybrid structure. (**d**) A high-magnified STEM image around morphotropic grain boundary. The atomic arrangements at two representative regions are present for clarification. (**e**) An atomic-resolved STEM image around grain boundary between (001)- and (110)-oriented LSCO layers,

demonstrating the atomically sharp interfaces with minimal dislocations and cation defects. (**f**) Intensity profiles obtained from line scans averaged across the respective regions in (**d**).

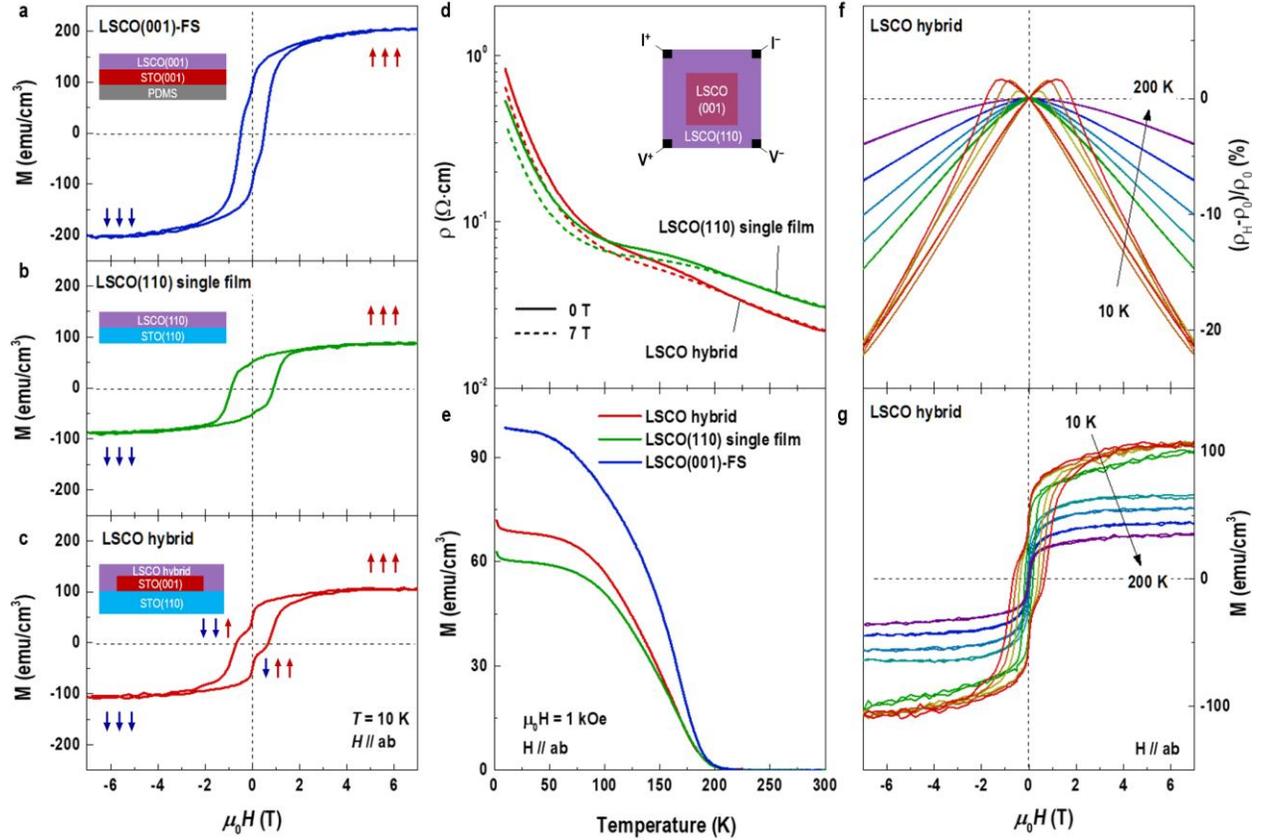

**Figure 2. Magnetometry and transport measurements on LSCO hybrid structures.** Field-dependent magnetization of (**a**) a (001)-oriented FS-LSCO membrane, (**b**) a (110)-oriented LSCO single film, and (**c**) a LSCO hybrid structure grown on (110)-STO substrates. The magnetic field was applied along the in-plane direction. *M-H* loops were measured at 10 K. Apparently, *M-H* loop of a LSCO hybrid structure is the superposition of those from a (001)-oriented FS-LSCO membrane and a (110)-oriented LSCO single layer. Temperature-dependent (**d**) resistivities (*ρ*) and (**e**) magnetization of a (110)-oriented LSCO single film and a LSCO hybrid structure on (110)-STO substrates. *ρ-T* curves were measured at 0 and 7 T. *M* versus *T* scans were collected during sample warming after field-cooling at 1 kOe. Field-dependent (**f**) magnetoresistance [($ρ_H−ρ_0$)/$ρ_0$] and (**g**) magnetization of a LSCO hybrid structure at various temperatures ranging from 10 to 200 K.

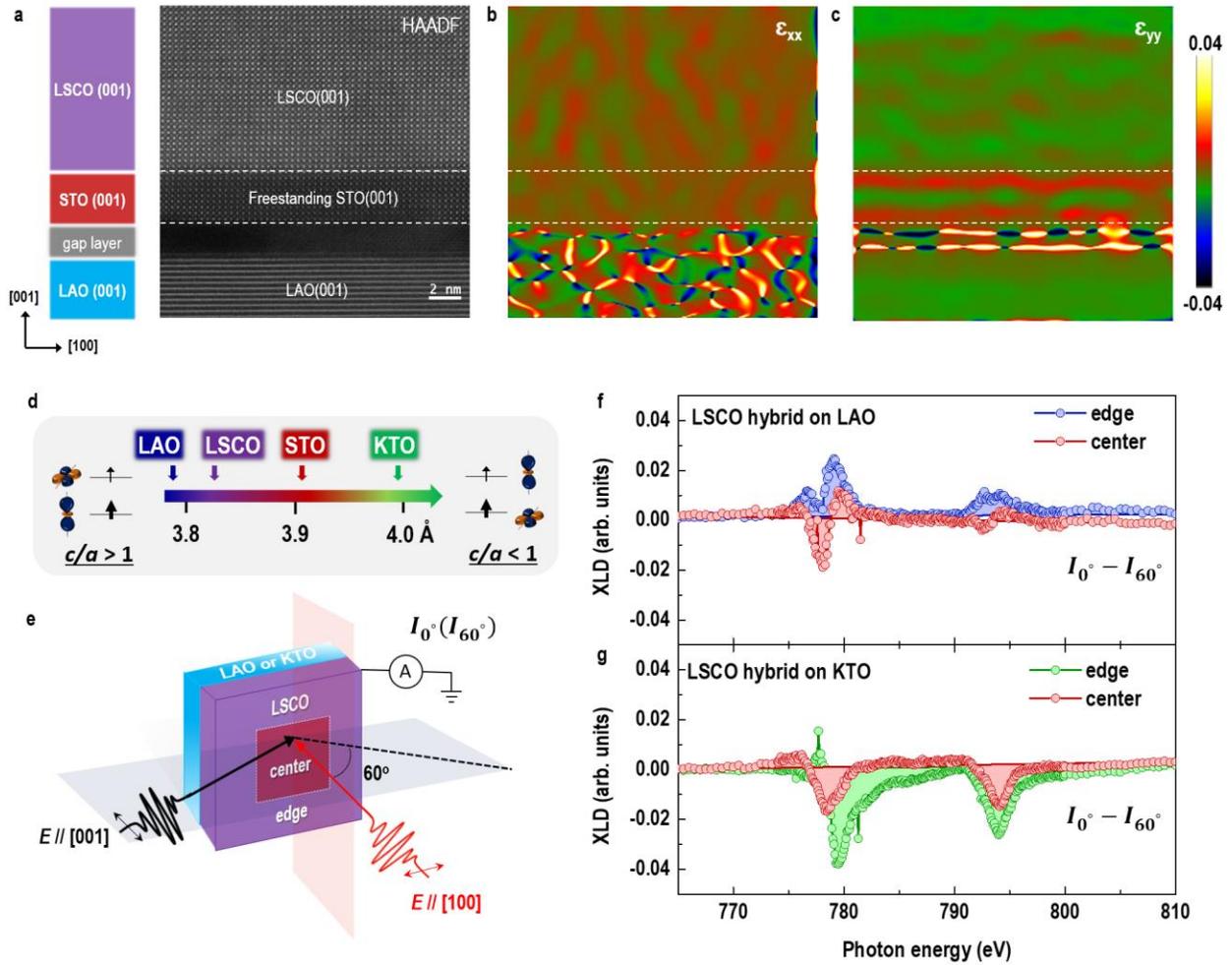

**Figure 3. Strain-mediated orbital polarization in LSCO hybrid structures.** (**a**) Schematic and HAADF-STEM image of a LSCO hybrid structure grown on (001)-oriented LAO substrates. (**b**) In-plane and (**c**) out-of-plane strain distributions within the LSCO hybrid structures performed by geometric phase analysis (GPA). (**d**) Lattice constants in angstroms of LSCO bulk and single crystalline substrates used in present work. The LSCO films grown on LAO is slightly compressed, whereas the LSCO is tensile-strained by STO and KTO substrates. (**e**) Schematic of the scattering geometry for XAS measurements with X-ray beam aligned parallel (0°) or with an angle of 60° respect to the surface normal. X-ray linear dichroism (XLD) spectra were quantified by calculating the difference between $I_{0°}$ and $I_{60°}$. (**f**) and (**g**) XLD of LSCO hybrids on LAO and KTO substrates, respectively. XLD spectra from the edge and center regions of LSCO hybrids show the distinct differences, indicating that the hole occupancy in 3*d* orbitals from these regions are in sharp contrast. The degree of orbital polarization exhibits a clear anisotropic orbital occupancy depending on the strain states of LSCO hybrids. All spectra were collected around the Co *L*-edges at 77 K in TEY mode.

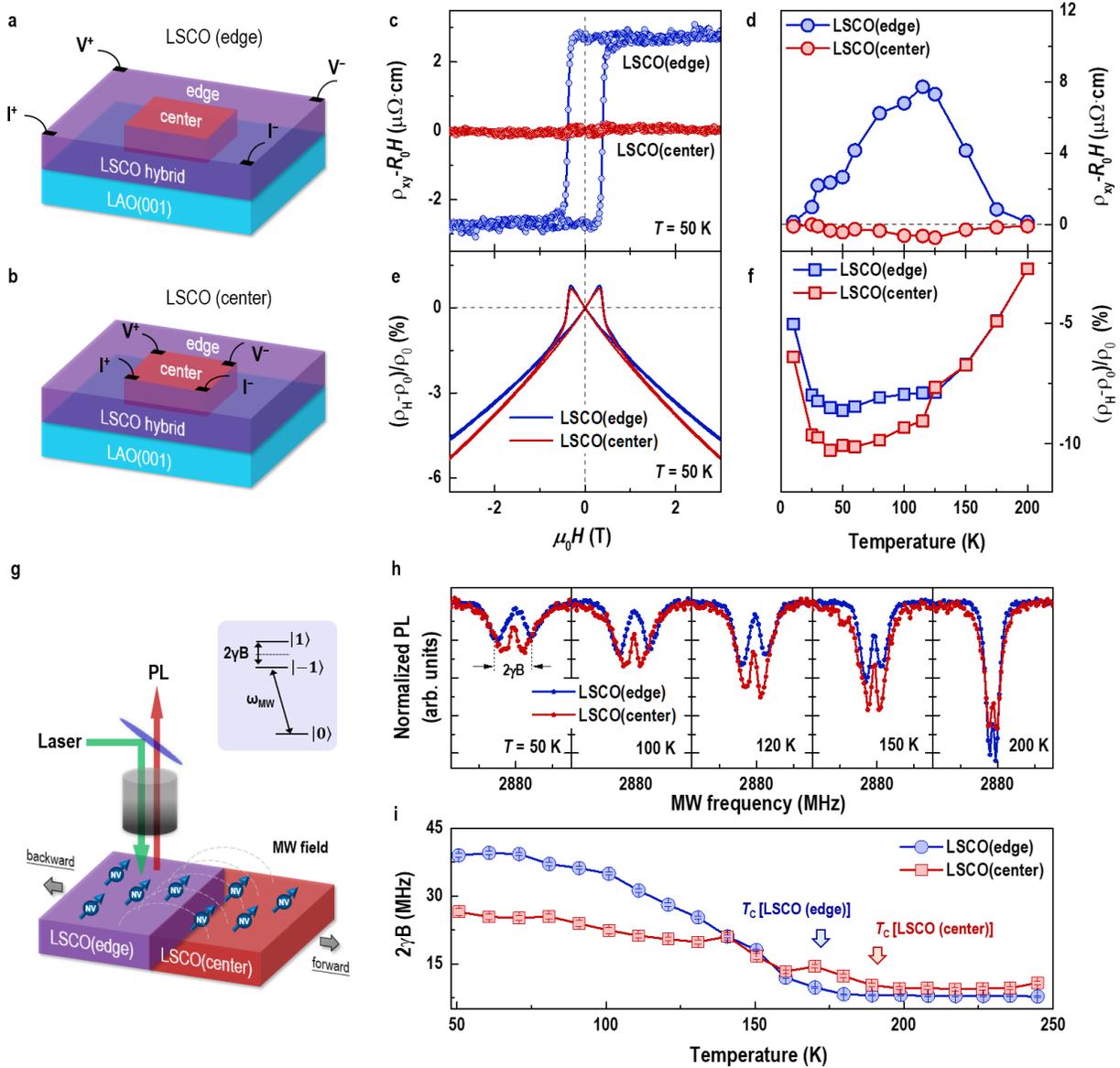

**Figure 4. Distinct ground states across grain boundary in LSCO hybrid structures.** (**a**) and (**b**) Schematic setups for transport measurements at the LSCO(edge) and LSCO(center) regions, respectively. Field-dependent (**c**) anomalous Hall resistance ($\rho_{xy}$-$R_0H$) and (**e**) magnetoresistance [($\rho_H$−$\rho_0$)/$\rho_0$] measured at the LSCO(edge) and LSCO(center) regions at 50 K. $R_0H$ represents the ordinary Hall term that is subtracted from $\rho_{xy}$ by linearly fitting the data at high magnetic field region. Temperature dependent (**d**) $\rho_{xy}$-$R_0H$ and (**f**) ($\rho_H$−$\rho_0$)/$\rho_0$ measured at the LSCO(edge) and LSCO(center) regions when $\mu_0H$ = 7 T. (**g**) Schematic of diamond NV-based magnetometry performed at different regions of LSCO hybrid structures grown on LAO substrates. (**h**) Zero-field optically detected magnetic resonance (ODMR) spectra of NV centers dispersed randomly on the surface of LSCO hybrid structures. The ODMR spectra were collected at different temperatures (Extended data fig. S18). (**i**) Temperature dependence of 2γB for LSCO(edge) and LSCO(center) regions, which exhibit different magnetic phase transition temperatures.

# Supplementary Materials for

## Braiding lateral morphotropic grain boundary in homogeneitic oxides

Shengru Chen,† Qinghua Zhang,† Dongke Rong,† Yue Xu, Jinfeng Zhang, Fangfang Pei, He Bai, Yan-Xing Shang, Shan Lin, Qiao Jin, Haitao Hong, Can Wang, Wensheng Yan, Haizhong Guo, Tao Zhu, Lin Gu, Yu Gong, Qian Li, Lingfei Wang, Gang-Qin Liu, Kui-juan Jin,* and Er-Jia Guo*

Correspondence to: kjjin@iphy.ac.cn and ejguo@iphy.ac.cn

**This PDF file includes:**

Figs. S1 to S19
References (*34–38*)

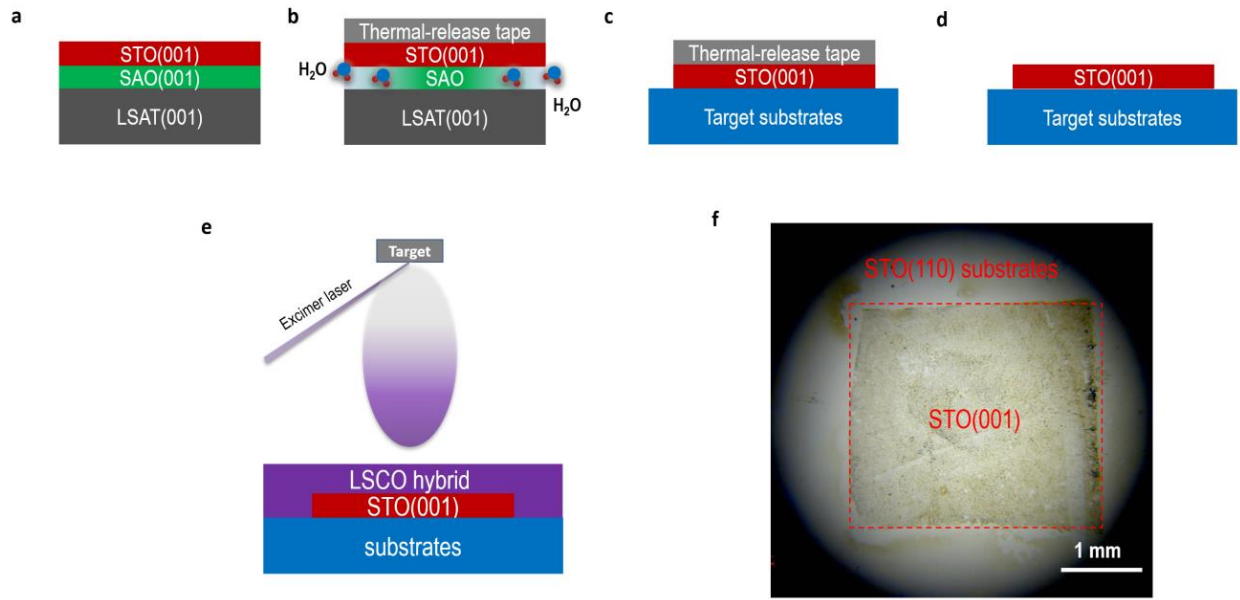

**Fig. S1. Synthesis of LSCO hybrid structures**. Process schematics for (a) (001)-oriented SAO/STO growth on (001)-oriented LSAT substrates, (b) water-solution of SAO layer and release ultrathin STO membranes, (c) transfer and (d) thermally release the STO membrane on different target substrates, for instance, the (110)- and (111)-oriented STO, LAO, and KTO substrates. (e) Fabrication of the LSCO hybrid structures on surface-engineered substrates by PLD technique. More experimental details see *Materials and Methods*. (f) Optical microscope image of an ultrathin STO membrane with a thickness of ~ 8-unit cells (u. c.) transferred onto the representative (110)-oriented STO substrates. The transferred STO membrane maintains its original shape and clear boundaries with minimal damage within the STO membranes. The typical size of the STO membranes is 3×3 mm$^2$.

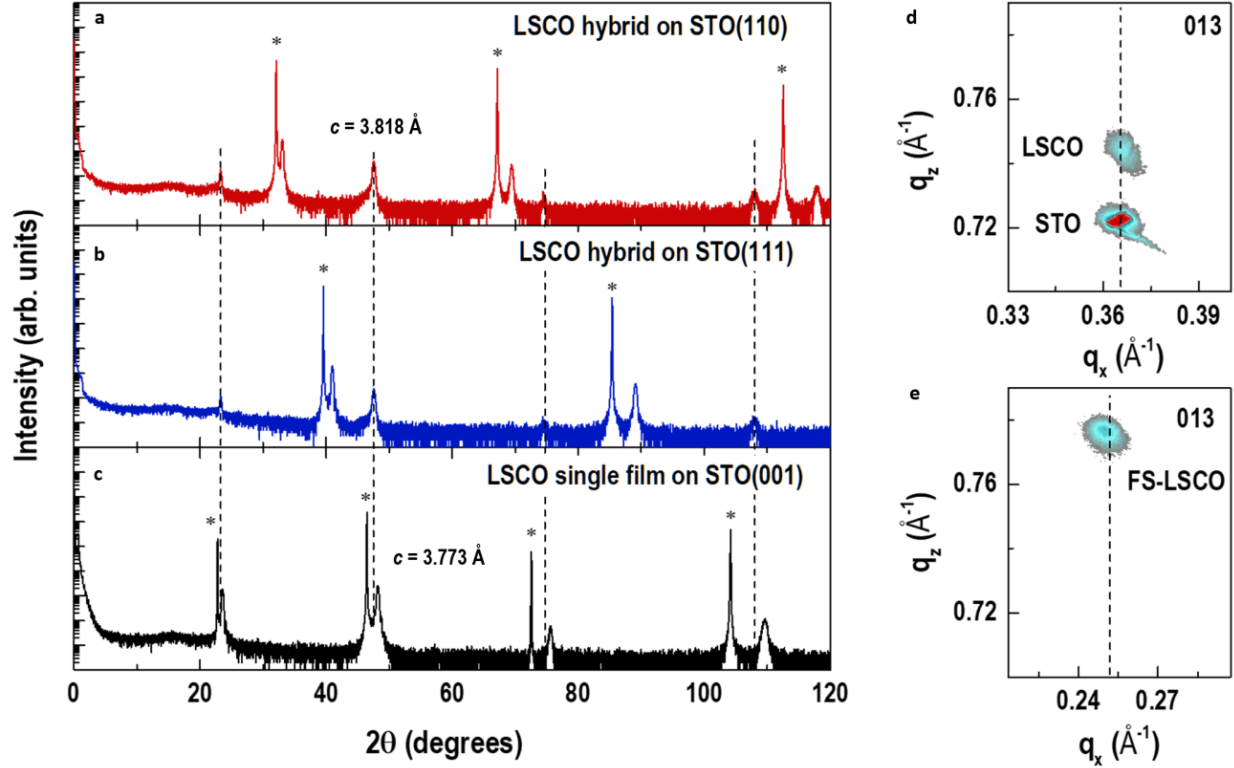

**Fig. S2. Epitaxial growth of LSCO hybrid structures**. XRD 2θ-ω scans of (a) a LSCO hybrid structure grown on a (110)-oriented STO substrate, (b) a LSCO hybrid structure grown on a (111)-oriented STO substrate, and (c) a (001)-oriented LSCO single film grown on STO substrates. The STO substrates' reflections are indicated with "*". The LSCO hybrids grown on (110)- and (111)-oriented STO substrates exhibit the clear (00$l$) reflections besides the typical reflections with identical orientations to the substrates. (d) Reciprocal space map (RSM) of a LSCO hybrid structure around 013 reflection of a (110)-oriented STO substrate. The results reveal that a 30 nm-thick LSCO film is coherently grown on the STO substrate. (e) RSM around 013 reflection of the (001)-oriented LSCO layers grown on the freestanding (FS) STO membranes. We find that the LSCO layers directly grown on STO substrates are tensile-strained. The out-of-lattice parameter ($c$) of LSCO layers is ~ 3.773 Å, whereas the $c$ of FS-LSCO layer grown on FS-STO membranes is ~ 3.818 Å. Please note that this FS-LSCO layer is not fully relaxed to its bulk form ($c_{pc}$ ~ 3.838 Å) but slightly tensile-strained by the FS-STO membranes. The macroscopic structural characterizations demonstrate the highly epitaxial growth of all LSCO layers on both STO substrates and (001)-oriented FS-STO membranes, forming a laterally LSCO hybrid structure with different crystallographic orientations.

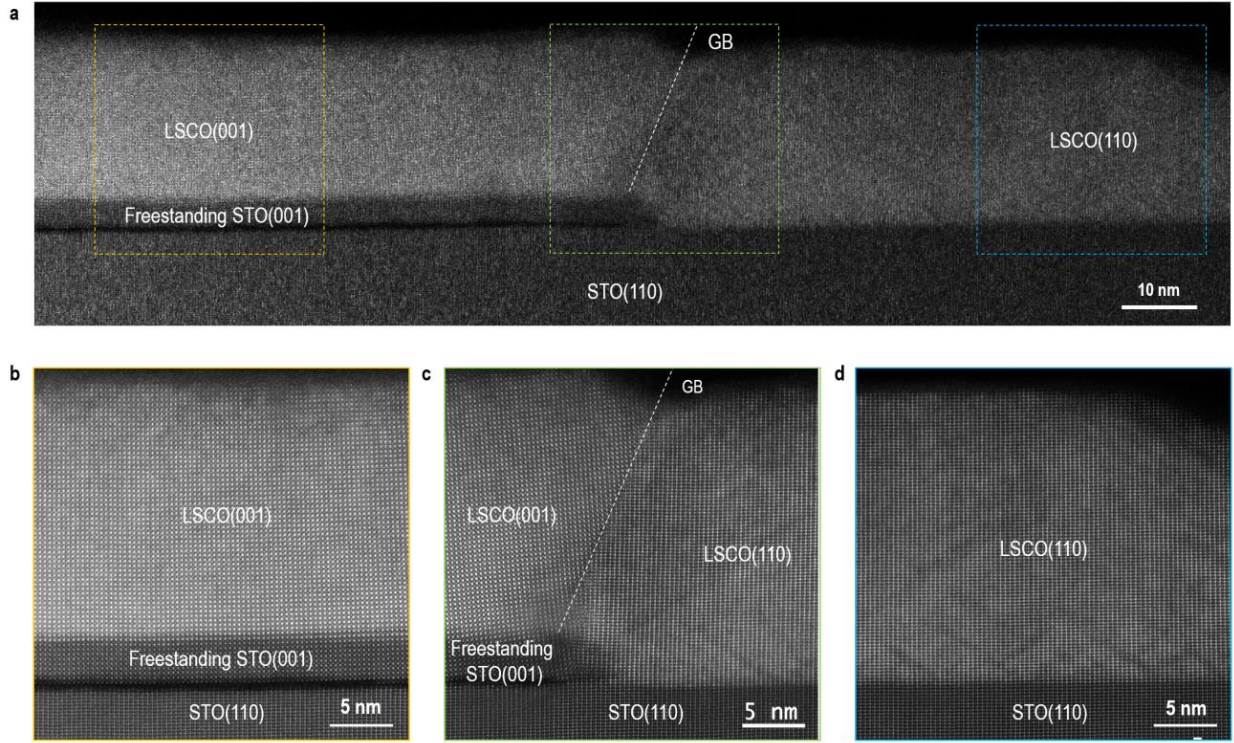

**Fig. S3. Microscopic structural characterization of a LSCO hybrid structure grown on (110)-oriented STO substrates**. (a) Low-magnified cross-sectional HAADF-STEM image of a LSCO hybrid structure. The FS-STO membrane is closely adhered to the (110)-oriented STO substrates and is continuous with homogenous layer thickness. High-magnified STEM images from the colored rectangles marked in (a) indicate three representative interface regions: (b) (001)-LSCO/(001)-FS-STO/(110)-STO region, (c) grain boundary (GB) region, and (d) (110)-LSCO/(110)-STO region. The brighter layers are LSCO, and the darker layers are STO, because the stronger electron scattering by heavier elements with larger atomic number (La > Sr). STEM results demonstrate that the interfaces between LSCO layers and STO substrates/membranes are atomically sharp, and the interface roughness is less than one-u. c.-thick. The white dashed lines indicate the positions of grain boundaries (GB) between different oriented LSCO layers. We note that there is a black region between the FS-STO membranes and (110)-oriented STO substrates. The gap is unbelievable narrow with an averaged thickness less than 1 nm. The naturally formed gap allows the FS-STO membrane unbonded to the (110)-oriented STO substrates, thus the LSCO layer follows the crystallographic orientation of FS-STO membranes.

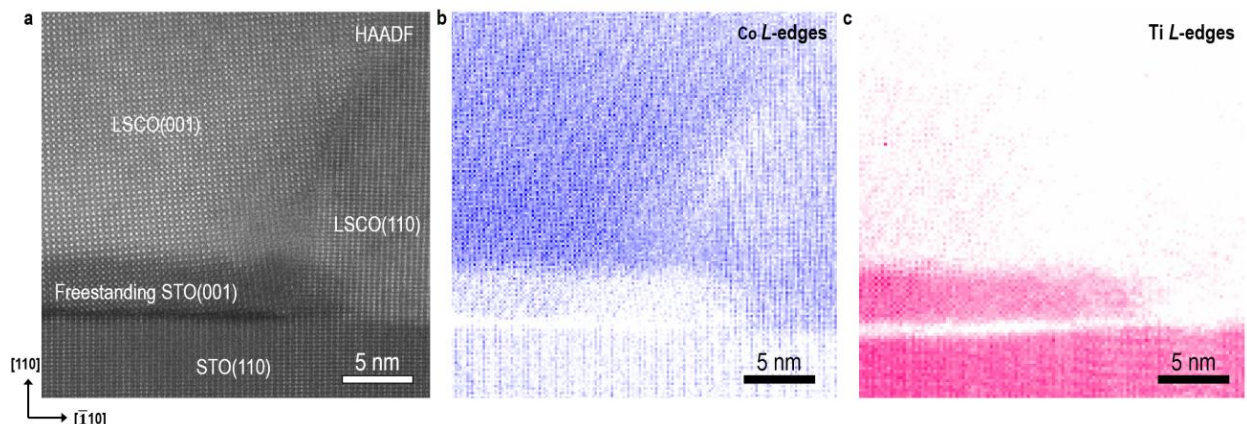

**Fig. S4. HAADF-STEM image and EELS map collected from a respective grain boundary region in LSCO hybrid structures**. (a) High-magnified STEM image around a grain boundary. The colored panels show the integrated EELS intensities of (b) Co $L_{3,2}$- and (c) Ti $L_{3,2}$-edges, respectively. We find that the interfaces between (001)-LSCO/(001)-FS-STO membranes and (110)-LSCO/(110)-STO substrates do not exhibit significant chemical intermixing. We determine that the chemical intermixing at both interfaces is approximately one-u. c.-thick.

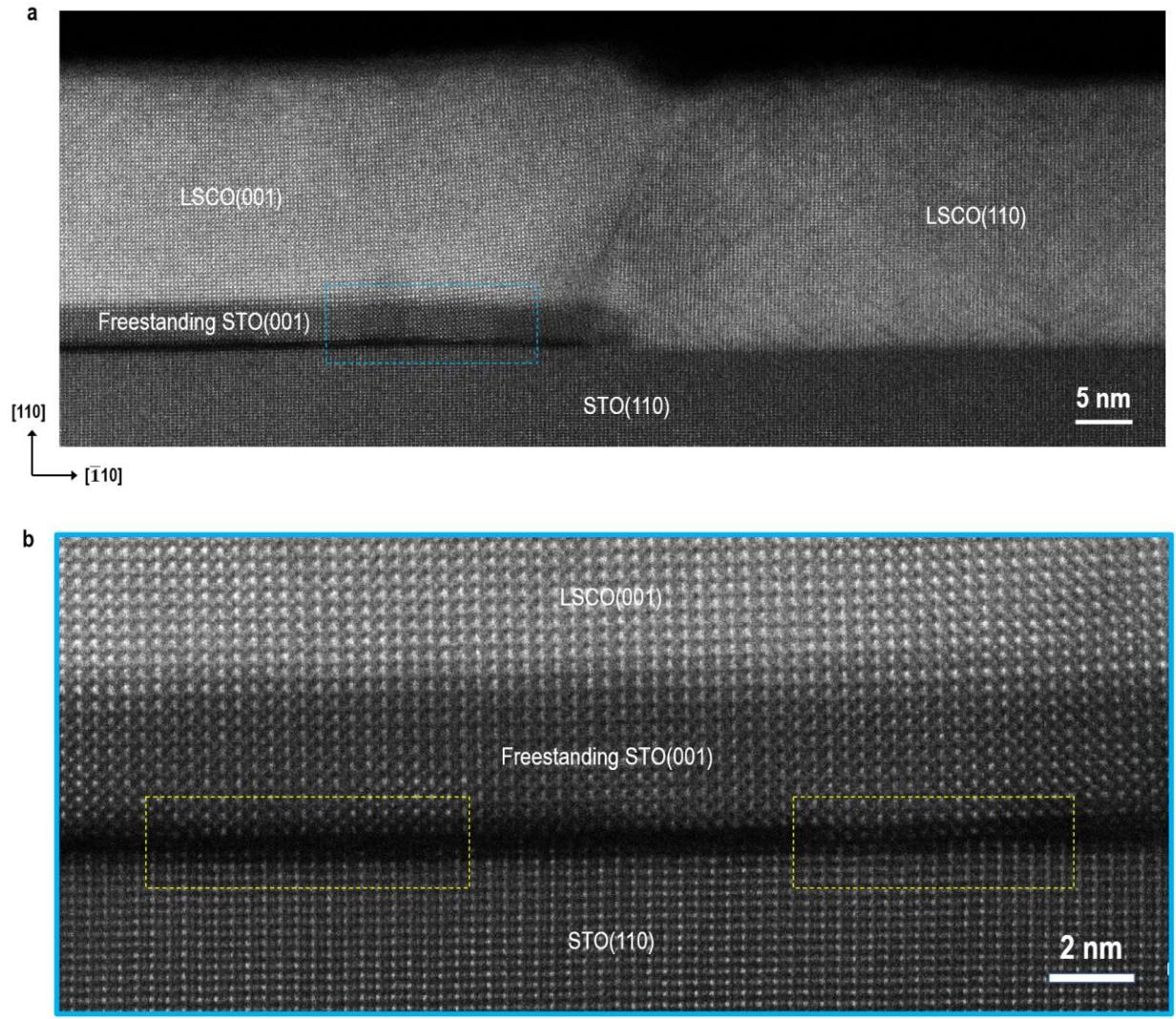

**Fig. S5. High-magnified HAADF-STEM image around interface region**. (a) Low-magnified STEM image of a LSCO hybrid structure grown on (110)-oriented STO substrates. (b) Atomic-resolved STEM image from a blue rectangle region in (a). The STEM results reveal that the averaged thickness of (001)-FS-STO membrane is ~ 8 u. c. (approximately 3 nm). The zoom-in STEM image illustrates that the thickness of gap is homogeneous and no more than 1 nm. The yellow dashed rectangle regions reveal that the possible formation of chemical bonding between the (001)-FS-STO membrane and (110)-STO substrates. We believe that the formation of these bonding at the interfaces can be attributed to the identical chemical composition and short distance between gangling bonds. However, please note that this chemical bonding is discontinuous and unstable due to the polar catastrophe (*37*). The (110) plane of STO substrate is charged, whereas the (001) plane of FS-STO is charge neutral. Thus, it prohibits forming the stable chemical bonding between (001)-FS-STO membranes and (110)-STO substates. The naturally formed gap at the interface region allows the FS-STO membranes to keep their own lattice parameters and crystallographic orientations, serving as independent templates for the epitaxial thin film growth.

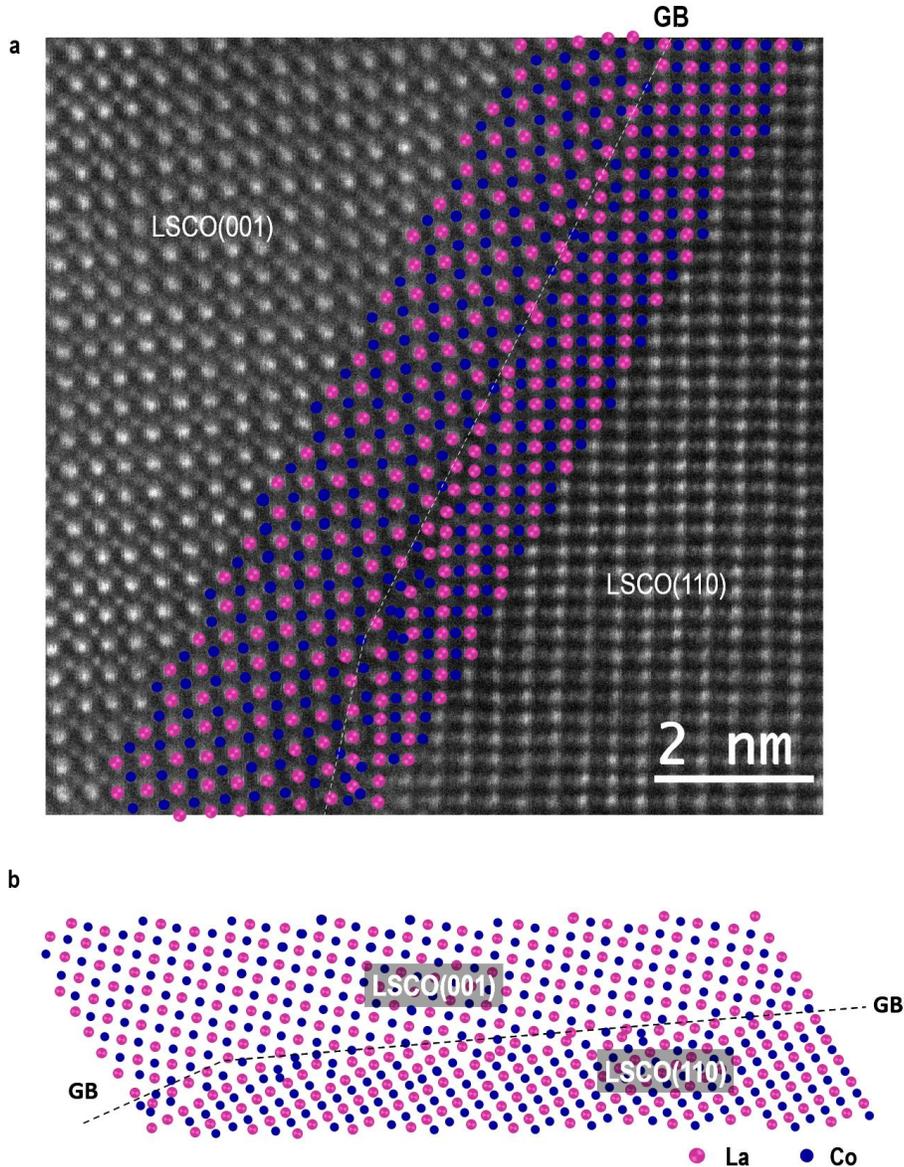

**Fig. S6. Atomic-resolved grain boundary in LSCO hybrid structures.** (a) HAADF-STEM image of a representative grain boundary (GB) between two LSCO domains with different orientations. The atom positions are acquired by fitting the intensity peaks with Gaussian function and are embed in the STEM image. The atomic structures around GB are shown in (b). The dashed lines represent the precise position of GB. We believe that the orientation of GB is possibly determined by minimizing misfit strain and thermally stabilization energy between LSCO domains with different orientations.

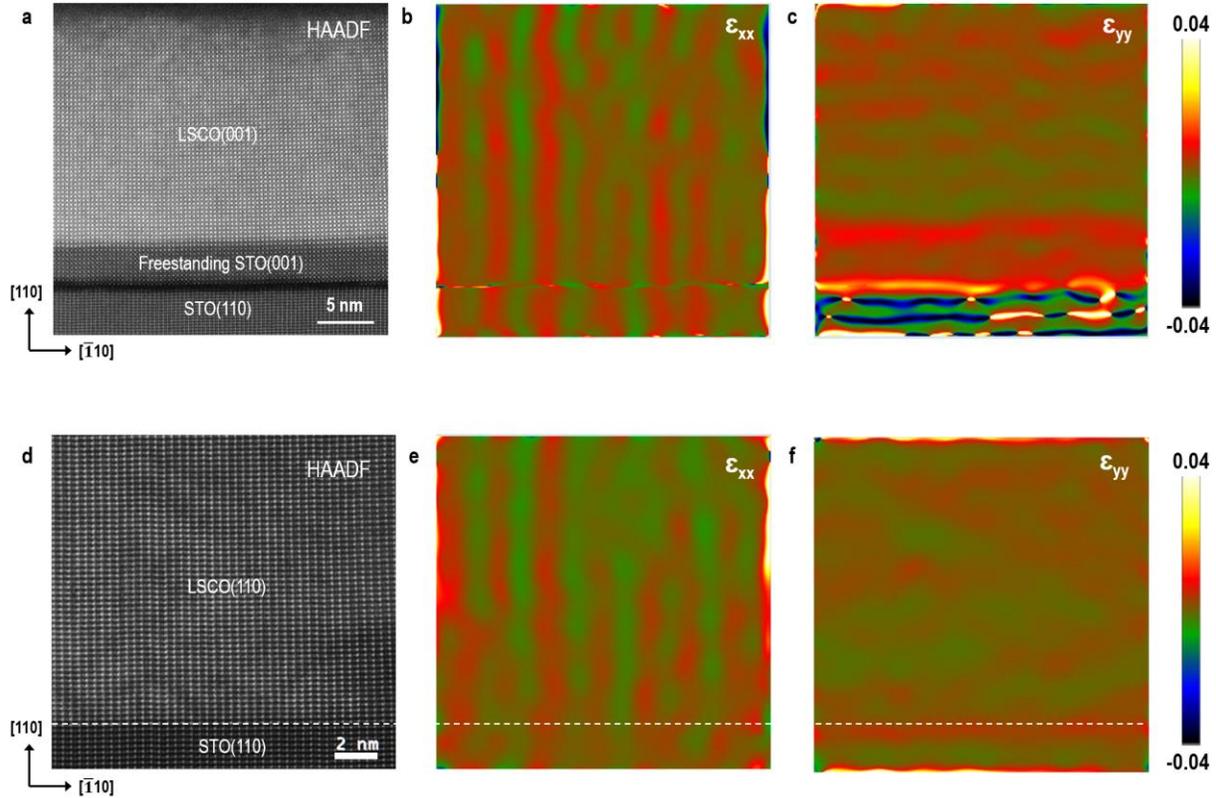

**Fig. S7. HAADF-STEM and geometric phase analysis (GPA) of LSCO hybrid structures.** (a) HAADF-STEM image of the (001)-LSCO/(001)-FS-STO/(110)-STO interface region. (b) In-plane and (c) out-of-plane strain distribution within the interface region in (a). The GPA results reveal that the (001)-LSCO layers possess nearly identical in-plane lattice constant of FS-STO membrane and (110)-STO substrates, demonstrating the as-grown (001)-LSCO layers still suffer from the in-plane tensile strain of FS-STO membrane, even though the thickness of membrane is only ~ 8 u. c. The out-of-plane lattice constant of (001)-LSCO layers is smaller than that of FS-STO membranes. (d) HAADF-STEM image of the (110)-LSCO/(110)-STO interface region. (e) In-plane and (f) out-of-plane strain distribution within the interface region in (d). The GPA results indicate that the (110)-LSCO layers are coherently grown on the (110)-STO substates and have the same in-plane lattice constants. The (110)-LSCO layers are tensile-strained, thus the out-of-plane lattice constant of films is smaller than that of (110)-STO substrates. However, the difference between the out-of-plane lattice constants of (110)-LSCO ($c_{LSCO110}$ ~ 2.68 Å) and (110)-STO ($c_{STO110}$ ~ 2.76 Å) is small. Therefore, the GPA is insensitive to such small variation when the LSCO films grown along the (110) orientation compared to the films grown along the (001) orientation.

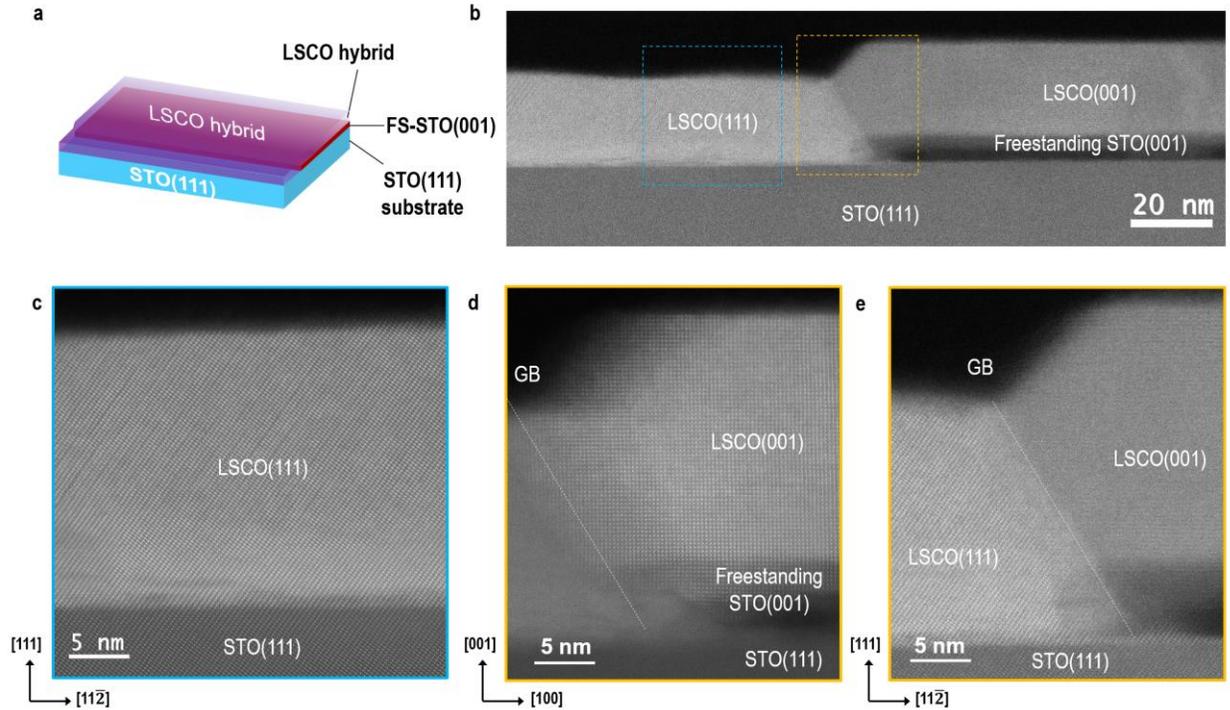

**Fig. S8. Microscopic structural characterization of LSCO hybrid structures on (111)-oriented STO substrates**. (a) Schematic of lateral homogeneitic LSCO hybrid structures grown on (111)-oriented STO substrates. (b) Low-magnified HAADF-STEM image of a LSCO hybrid structure grown on (111)-oriented STO substrates. The dark layer is the (001)-oriented FS-STO membranes, which provides a growth template for the (001)-oriented LSCO layers. (c) Cross-sectional STEM image of (111)-LSCO/(111)-STO interface region. The interface between LSCO layers and STO substrates is atomically sharp. The (111)-oriented LSCO layers are structurally uniform. Cross-sectional STEM images of grain boundary (GB) region in a LSCO hybrid structure are shown in (d) and (e). The (001)-FS-STO membrane is twisted unconsciously in a small angle with respect to the STO substrates, i. e. the [100] orientation of FS-STO membranes is not perfectly aligned to the [11$\bar{2}$] or [$\bar{1}$10] orientations of (111)-STO substrates. The twisted structures can be arbitrarily controlled through the alignment process during the FS-STO transfer onto the substrates. The imperfect alignment between FS-STO membranes and (111)-STO substrates prevents us acquiring the atomic-resolved STEM images from both interface regions simultaneously. The twisted angle is determined to be ~ 10 degrees by rotating the sample stick stepwise. The STEM image indicates that the uniformed structure within the (001)-LSCO layers and the (001)-LSCO/(001)-FS-STO interface is atomically sharp. From (e), the GB between (001)- and (111)-oriented LSCO layers can be identified clearly. The orientation of GB is determined by the minimum thermally stabilized energy between two different oriented LSCO domains.

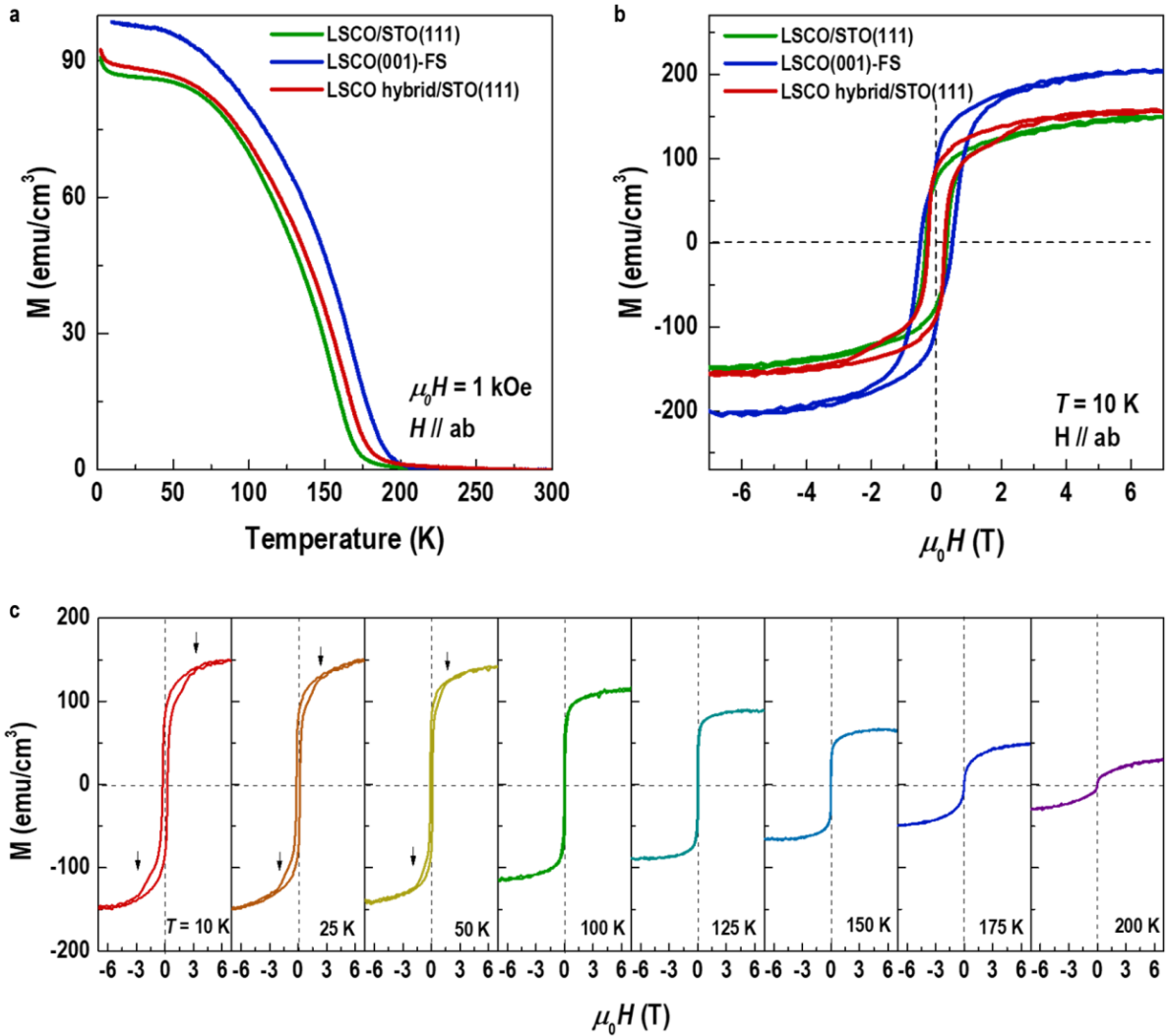

**Fig. S9. Magnetic measurements on a LSCO hybrid structure on (111)-oriented STO substates**. (a) Temperature dependent magnetization of a (111)-LSCO single film, a (001)-LSCO/STO freestanding membrane, and a LSCO hybrids on (111)-STO substrates. *M-T* curves were measured during the sample warming-up after cooled at an in-plane magnetic field of 1 kOe. The $T_C$ of a (001)-LSCO/STO freestanding membrane is ~ 200 K, followed by $T_C$ of a LSCO hybrids on (111)-STO substrates is ~ 183 K and $T_C$ of a (111)-LSCO single film is ~ 175 K. (b) Field dependent magnetization of a (111)-LSCO single film, a (001)-LSCO/STO freestanding membrane, and a LSCO hybrids on (111)-STO substrates at 10 K. *M-H* loop of a LSCO hybrid structure contains both features in the hysteresis loops of a (111)-LSCO single film and a (001)-LSCO/STO freestanding membrane, suggesting that the contribution to the magnetization of LSCO hybrids is twofold. *M-H* loops of a LSCO hybrids measured at different temperatures are present in (c). The black arrows indicate the typical feature from a (001)-LSCO/STO freestanding membrane. The critical magnetic field decays gradually as increasing temperature, suggesting the reduction of coercive field of a (001)-LSCO/STO freestanding membrane. When *T* beyond 100 K, the saturation magnetization of a LSCO hybrid structure decays rapidly.

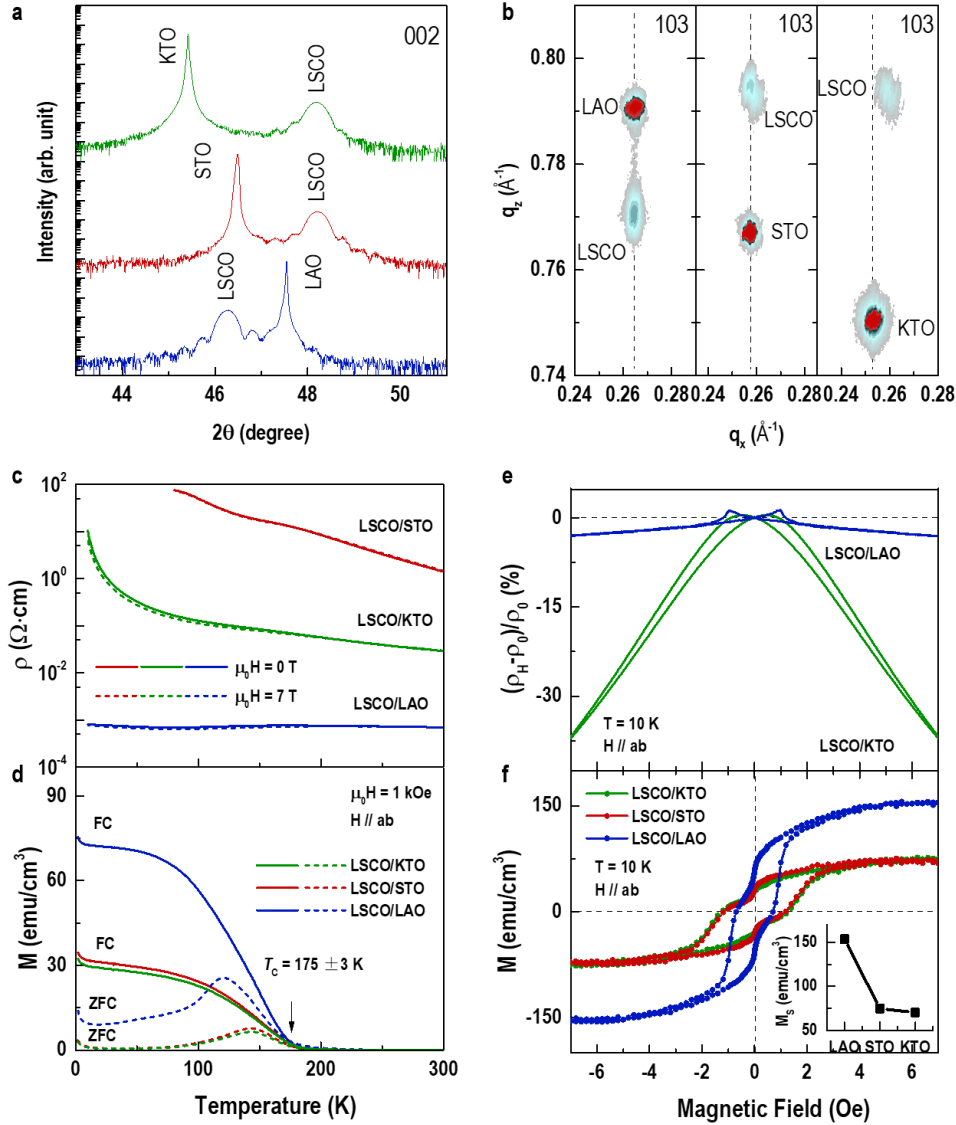

**Fig. S10. Strain-mediated physical properties of LSCO thin films**. (a) XRD 2θ-ω scans around 002 reflections of 30-nm-thick LSCO single films grown on LAO, STO, and KTO substrates. Clear Laue oscillations suggest that all LSCO films are epitaxially grown with high crystallinity. (b) RSMs around 103 reflections of strained LSCO single films. The LSCO films are coherently strained by LAO and STO substrates, yielding to the out-of-plane lattice parameters (*c*) of LSCO films are 3.922 and 3.775 Å, respectively. The *a* and *c* of LSCO films on KTO is 3.891 Å and 3.773 Å, respectively, demonstrating the strain relaxation of LSCO films. Temperature dependent (c) resistivity (*ρ*) and (d) magnetization (*M*) of LSCO films reveal the ferromagnetic character and spin-glass behavior at low temperatures. The LSCO films under tensile strains exhibit an insulating behavior, whereas the compressive-strained LSCO film undergoes an insulator-to-metal transition around $T_C$ and shows a resistivity upturn at ~100 K. (e) Magnetoresistance (MR) and (f) *M* of LSCO films as a function of magnetic field at 10 K. MR of LSCO films on KTO is an order of magnitude larger than that of LSCO layers on LAO. We observe a strong reduction of ordered magnetic moment with growing in-plane strain, in consistent with early work (*38*).

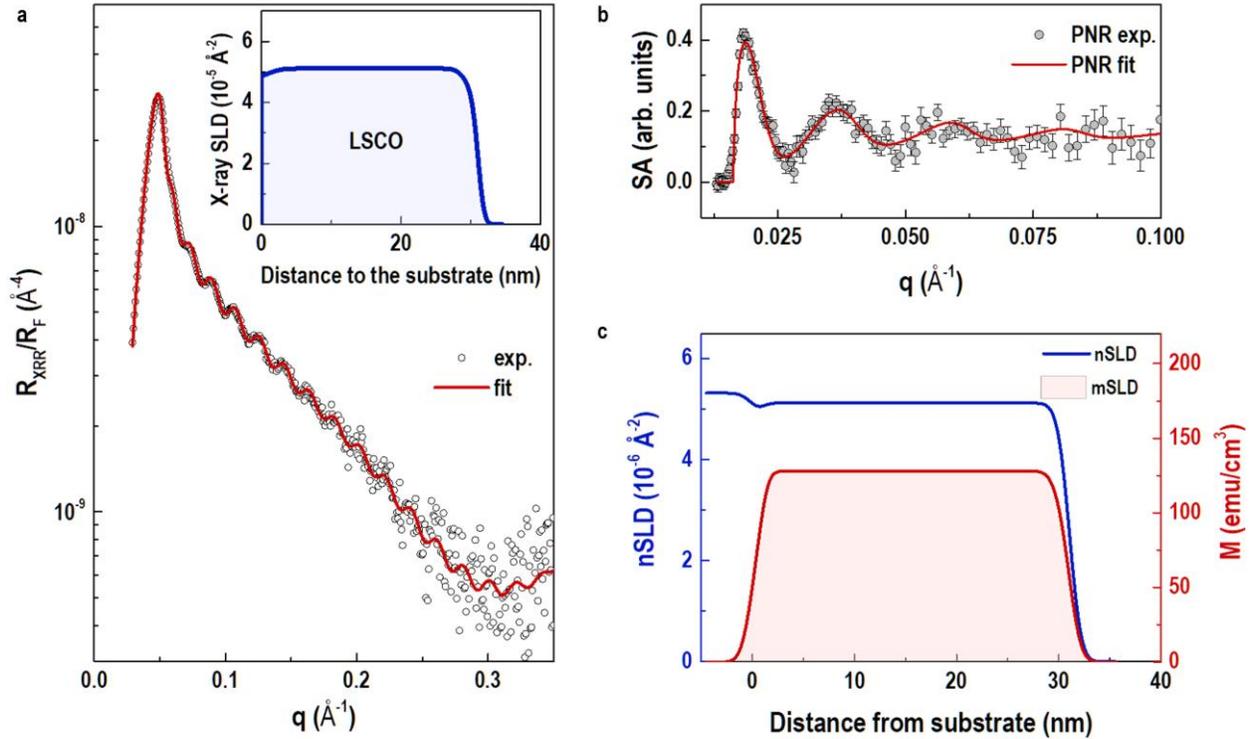

**Fig. S11. Uniformly distributed magnetization within strained LSCO films probed by PNR.**
(a) XRR curve of a LSCO film grown on a LAO substrate. The solid line is the best fitting to the experimental data (open circles). The thickness of LSCO film is 30.6 ± 0.4 nm and the surface/interface roughness is ~ 5 Å. Inset of (a) shows the X-ray scattering length density (SLD) depth profile of a LSCO film grown on a LAO substrate. The chemical composition and structural parameters are used for the polarized neutron reflectivity (PNR) fitting. (b) Spin asymmetry (SA) as function of wave vector $q$ (= $4\pi\sin\theta/\lambda$), where θ is the incident angle and λ is the wavelength of neutron beam. The SA is derived from the different neutron reflectivities for spin-up ($R^+$) and spin-down ($R^-$) polarized neutrons and calculated by $(R^+−R^-)/(R^++R^-)$. The solid red line is the best fit to the experimental data. (c) Magnetization ($M$) and nuclear SLD (nSLD) depth profile of a LSCO film grown on a LAO substrate. The magnetization (~ 123 emu/cm$^3$) is uniformly distributed within the LSCO film, except for a slightly reduced magnetization at the surface and interface due to the different boundary conditions. This result agrees well with the magnetometry result (~110 emu/cm$^3$ at 1 T) on LSCO films grown on LAO substrates.

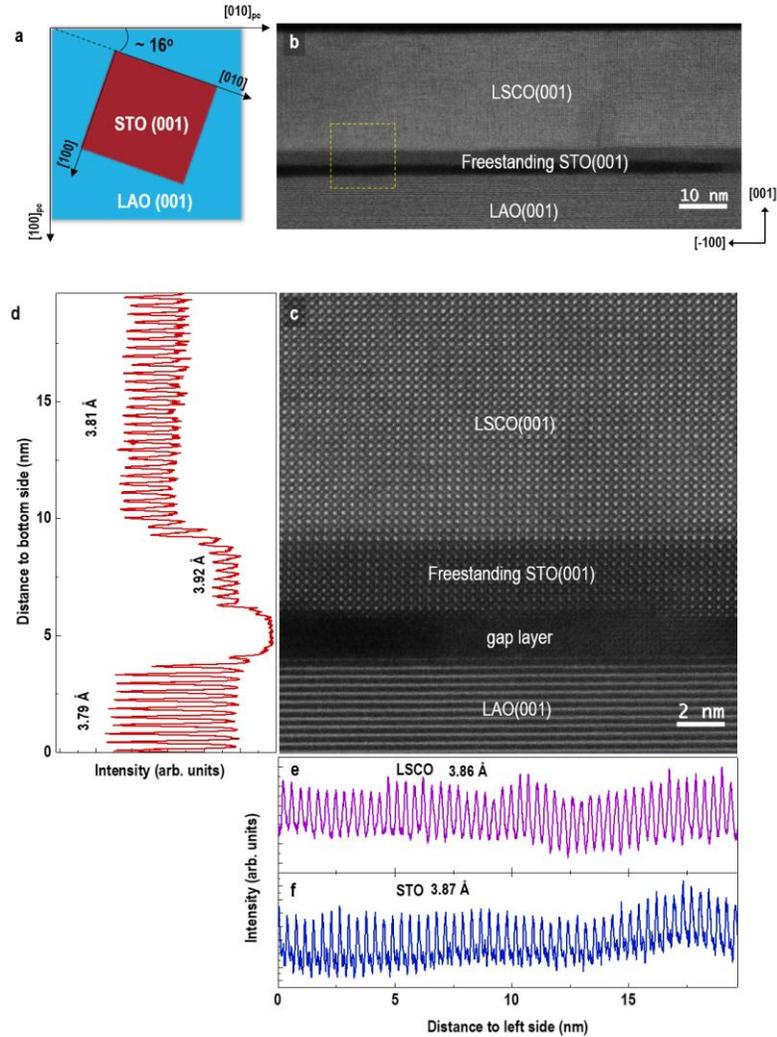

**Fig. S12. Structural analysis of LSCO hybrid structures on LAO substrates**. (a) Schematic of twisted ultrathin FS-STO membranes transferred on (001)-oriented LAO substrates. The twisted angle between [010] orientation of FS-STO membranes and $[010]_{pc}$ orientation of LAO substrates is ~ 16°, determined by sample rotation in the microscopy (b) HAADF-STEM image of LSCO hybrid structures on LAO substrates. The brighter regions in the STEM image are LAO and LSCO due to the stronger electron scattering from heavier elements (La). An atomic-resolution STEM image at the interface region is present in (c). We plot the intensity line profiles along (d) the out-of-plane direction, (e) LSCO film region, and (f) STO membrane region. The averaged in-plane lattice parameter of (001)-oriented LSCO layer is ~ 3.86 Å, which is slightly smaller than that of FS-STO membrane (~ 3.87 Å). Although the in-plane strain is slightly relaxed, the LSCO layers still suffer from the tensile strain of STO membranes. This case is different from a LSCO single film with identical thickness coherently strained to the STO single crystal substrates. The thickness of STO membranes (~ 8 u. c.) is an order of magnitude smaller than the thickness of LSCO layers (~ 80 u. c.). The lattice structure of STO membranes may deforms slightly to comply the elastic misfit strain between STO and LSCO. A small lattice elongation of FS-STO membrane is observed along the out-of-plane direction. In addition, we note that the gap between FS-STO and LAO substrates is ~ 2 nm in thickness. There is no apparent chemical bonding between FS-STO and LAO.

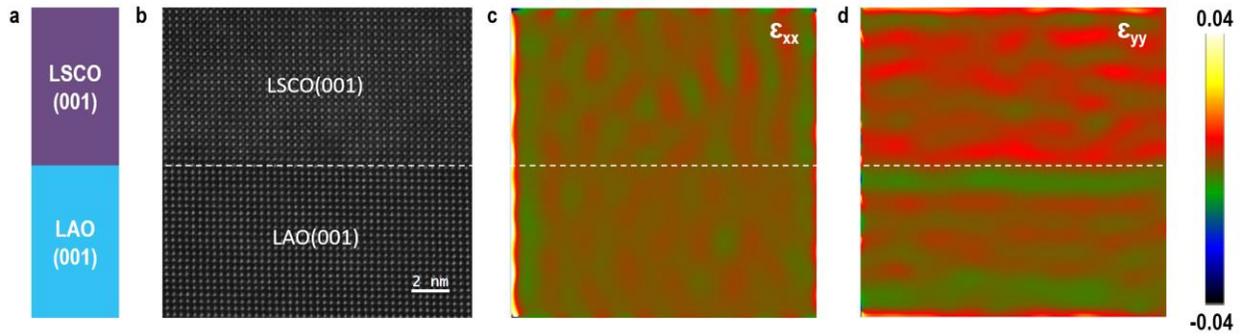

**Fig. S13. Strain distribution within an LSCO layer grown directly on LAO substrates**. (a) Schematic of sample configuration. (b) A HAADF-STEM image of the interface region between LSCO layers and LAO substrates. Since the heavy elements in LSCO and LAO are identical (La), it is difficult to distinguish the interfaces between those two materials. We perform the geometric phase analysis (GPA) at the same region. (c) and (d) show the in-plane and out-of-plane strain distributions at the interface region, respectively. The LSCO layers are coherently compressive-strained to LAO substrates. The in-plane lattice constant of LSCO layers is identical to that of LAO substrates. Thus, the LSCO layers is elongated along the out-of-plane direction. We determine that the out-of-plane lattice constant of LSCO layers is approximately 3.92 Å, which is significantly larger than that of LAO substrates (~ 3.79 Å).

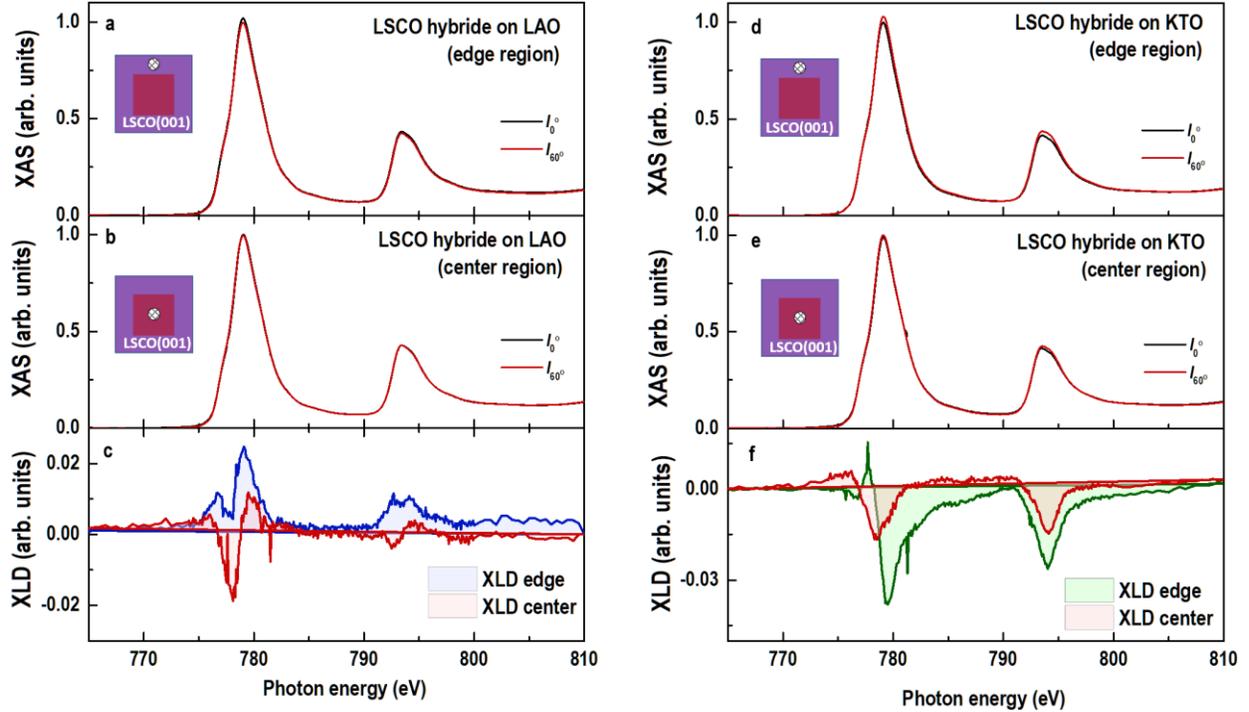

**Fig. S14. Regional dependent electronic states of LSCO hybrid structures**. (a) and (b) XAS at Co $L_{3,2}$-edges measured at the edge and center regions of LSCO hybrid structures on LAO substrates, respectively. The edge region of LSCO hybrid structure is the direct growth of (001)-LSCO layers on (001)-oriented LAO substrates. The LSCO layers is compressively strained, similar to the LSCO single films grown on the LAO substrates. The center region of the LSCO hybrid structure possesses the (001)-LSCO layers grown on the FS-STO membranes. In this case, the (001)-LSCO layers at the center region is slightly tensile-strained to the FS-STO membranes. The distinct strain states of LSCO layers determine the opposite XLD at different regions [(c)]. In the center region of LSCO hybrid structure, the electrons prefer occupying the $d_{x^2-y^2}$ orbitals, whereas the electron occupancy in the $d_{3z^2-r^2}$ orbitals is larger at the edge region of LSCO hybrid structure. The control experiments were conducted on a LSCO hybrid structure grown on KTO substrates [(d) and (e)]. Both FS-STO membranes and KTO substrates apply the in-plane tensile strain to the LSCO layers. The signs of XLD are the same for the edge and center regions of LSCO hybrid structures. The amplitude of XLD at the edge region (LSCO/KTO) is larger than that of XLD measured at the center region ((001-)LSCO/FS-STO) because the LSCO layers suffer a larger tensile strain from the KTO substrates. Thus, more electrons occupy the $d_{x^2-y^2}$ orbitals at the edge regions compared to the center regions.

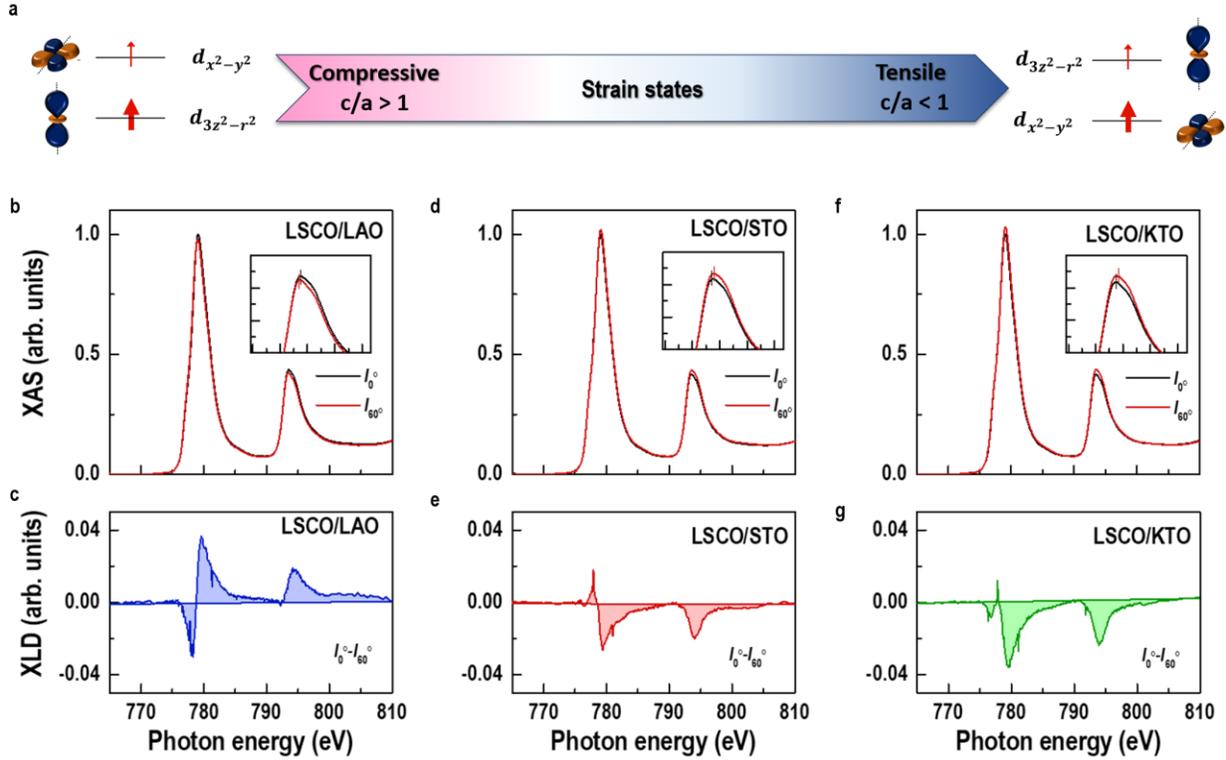

**Fig. S15. Electronic states of strained LSCO single films**. (a) Schematic of electronic states under different strain states. The X-ray beam with linearly polarization incidents on the sample surface. The X-ray beam is parallel to the surface normal or has an incident angle of 60º with respect to the surface normal. The electronic state of Co *3d* orbitals is probed by measuring the resonance X-ray absorption spectra. When the X-ray beam is parallel to the surface normal, the XAS ($I_{0°}$) reflects the orbital occupancy of Co $d_{x^2-y^2}$ orbital directly. When the X-ray beam incidents on the sample surface with an angle of 60º, the XAS ($I_{60°}$) contains both orbital information from Co $d_{x^2-y^2}$ and Co $d_{3z^2-r^2}$ orbitals. (b), (d), and (f) XAS at Co *$L_{3,2}$*-edges for LSCO films grown on (001)-oriented LAO, STO, and KTO substrates, respectively. Apparently, the XAS signals from $Co^{3+}$ ions dominate the entire XAS spectra due to the 80% Co ions is +3 and only 20% Co ions is +4. The black and red curves represent the XAS ($I_{0°}$) and XAS ($I_{60°}$), respectively. In LSCO/LAO, the intensity and peak energy of XAS ($I_{60°}$) are smaller than those of XAS ($I_{0°}$), indicating that the most of holes occupy the $d_{x^2-y^2}$ orbitals and the free electrons occupy the $d_{3z^2-r^2}$ orbitals. By contrast, the electrons will occupy the $d_{x^2-y^2}$ orbitals when the LSCO films are tensile-strained. (c), (e), and (g) X-ray linear dichroism (XLD) for the LSCO films grown on (001)-oriented LAO, STO, and KTO substrates, respectively. The XLDs are simply calculated by $I_{0°} - I_{60°}$. The orbital polarization of LSCO films varies with the strain states. XLD results reinforce the predominant $d_{x^2-y^2}$ (or $d_{3z^2-r^2}$) hole character in compressively (or tensile) strained LSCO films.

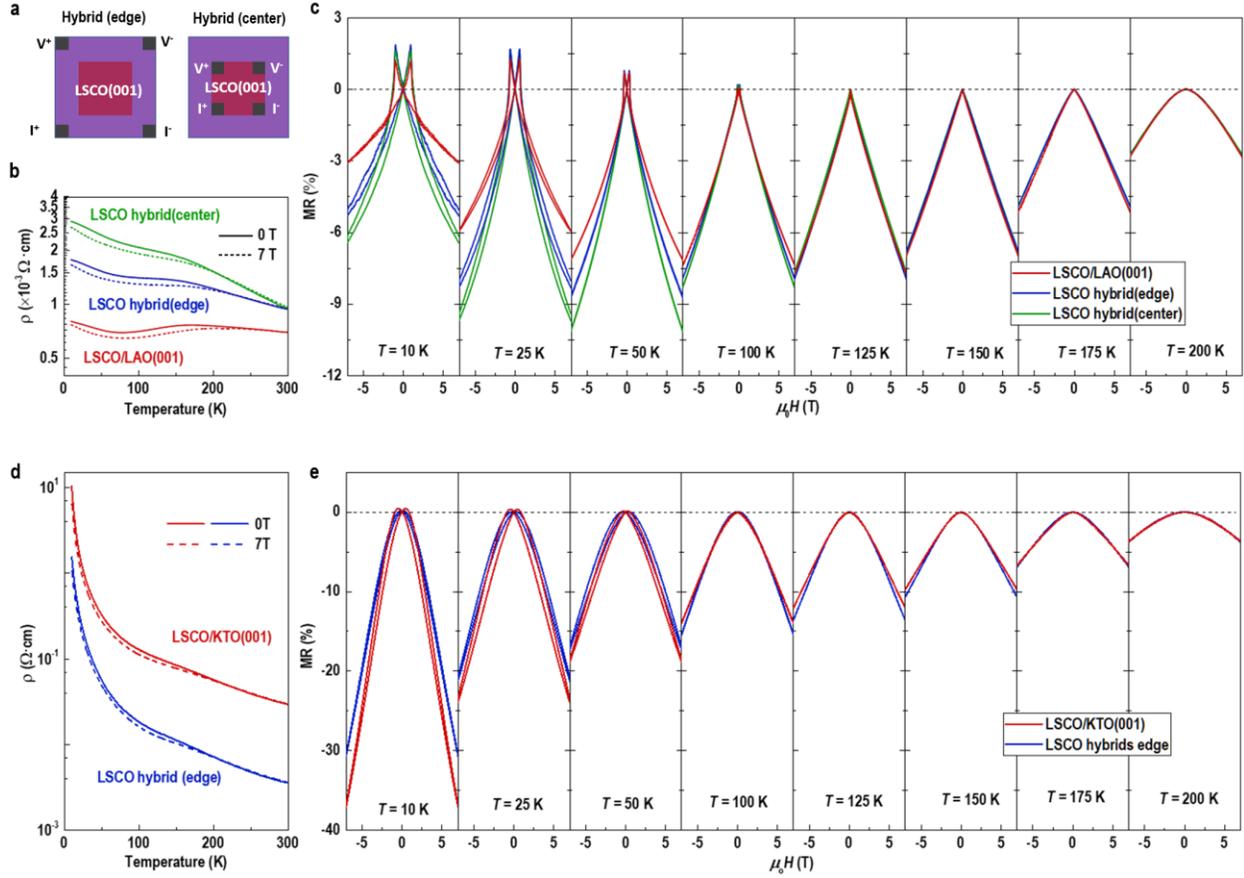

**Fig. S16. Transport measurements of LSCO hybrid structures.** (a) Schematic setups for transport measurements at edge and center regions of LSCO hybrid structures. (b) Temperature dependent resistivity ($\rho$) of a (001)-oriented LSCO single film, the edge and center regions of LSCO hybrid structures on LAO substrates. The transport measurements were performed in van der Paw method with applying the out-of-plane magnetic fields. The $\rho$ of LSCO hybrid structures is larger than that of a (001)-oriented LSCO single film grown on LAO because the LSCO layers under tensile strain states exhibit an insulating behavior in contrast to the metallic conductivity of a compressively strained LSCO layers. (c) Magnetoresistance [MR = $(\rho_{7T}-\rho_{0T})\times 100\%/\rho_{0T}$] of a (001)-oriented LSCO single film, the edge and center regions of LSCO hybrid structures on LAO substrates at different temperatures. The MR of LSCO hybrid structures is a few times larger than that of a LSCO film grown on LAO substrates at low temperatures. This fact agrees well with the reduction of ordered magnetic domains as increasing in-plane tensile strain. The difference between MRs of (001)-LSCO/LAO and LSCO hybrid structures becomes small at high temperatures. The physical picture is further reinforced by performing transport measurements on a LSCO hybrid structure grown on KTO substrates [(d) and (e)]. The $\rho$ of LSCO hybrid structures reduces by less than an order of magnitude compared to that of a LSCO single film grown on KTO substrates. Meanwhile, the MR of LSCO hybrid structure is slightly smaller than that of a LSCO single film on KTO. These results support the strain-driven suppression of the ferromagnetic phase in LSCO layers.

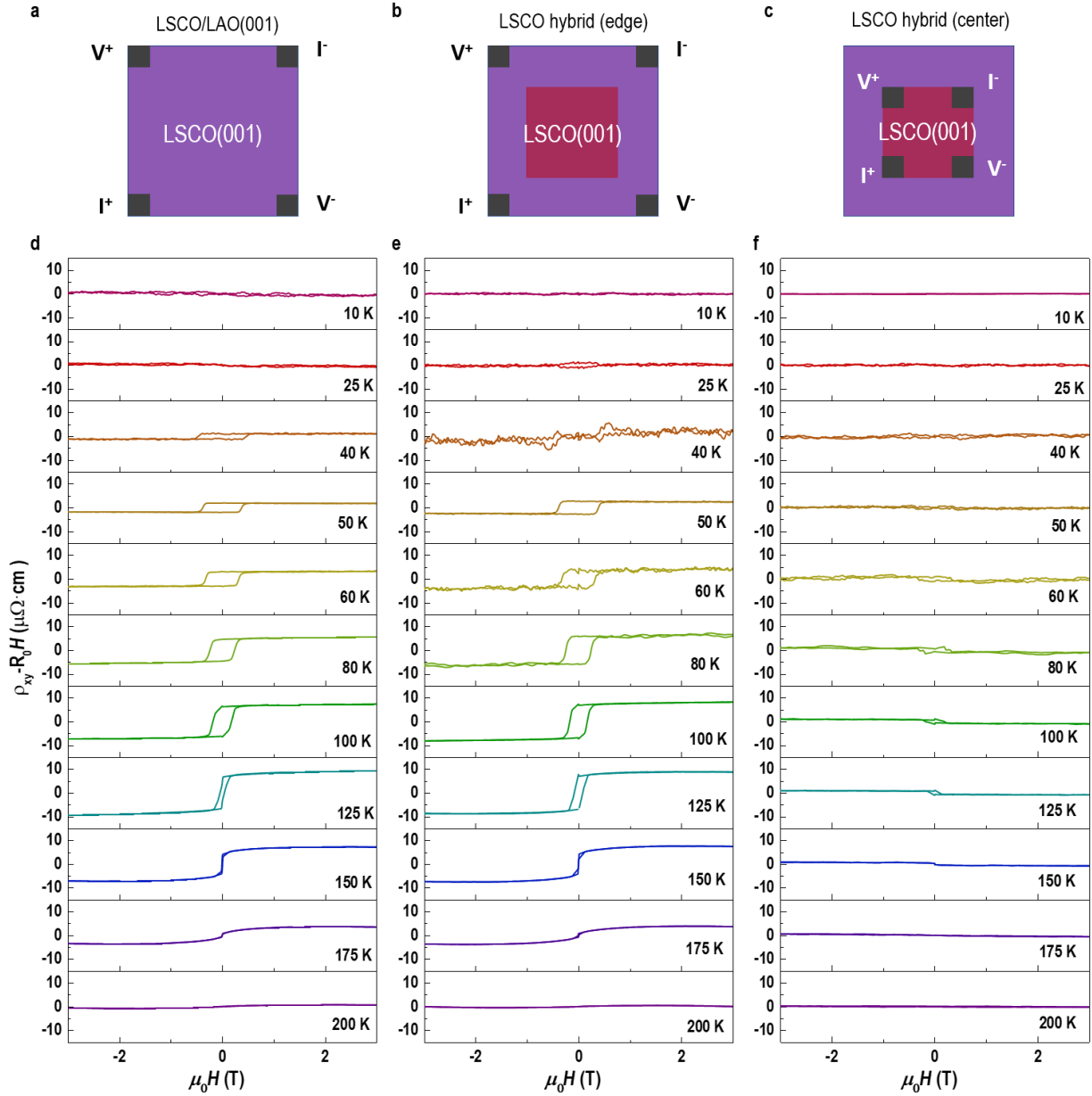

**Fig. S17. Anomalous Hall resistivities of LSCO single films and hybrid structures**.
Schematic measurement configurations for (a) LSCO single films grown on (001)-LAO substrates (LSCO/LAO), (b) the edge region and (c) center region of LSCO hybrid structure grown on LAO substrates. (d), (e), and (f) Field-dependent Hall conductivities at different temperatures for a LSCO/LAO, the edge and center regions of LSCO hybrid structures, respectively. The $R_0H$ is subtracted from $\rho_{xy}$ by linearly fitting the data at high magnetic field region. The ($\rho_{xy}-R_0H$) at 7 T for the LSCO hybrid (edge) and LSCO hybrid (center) are summarized in Figs. 4C and 4E of main text.

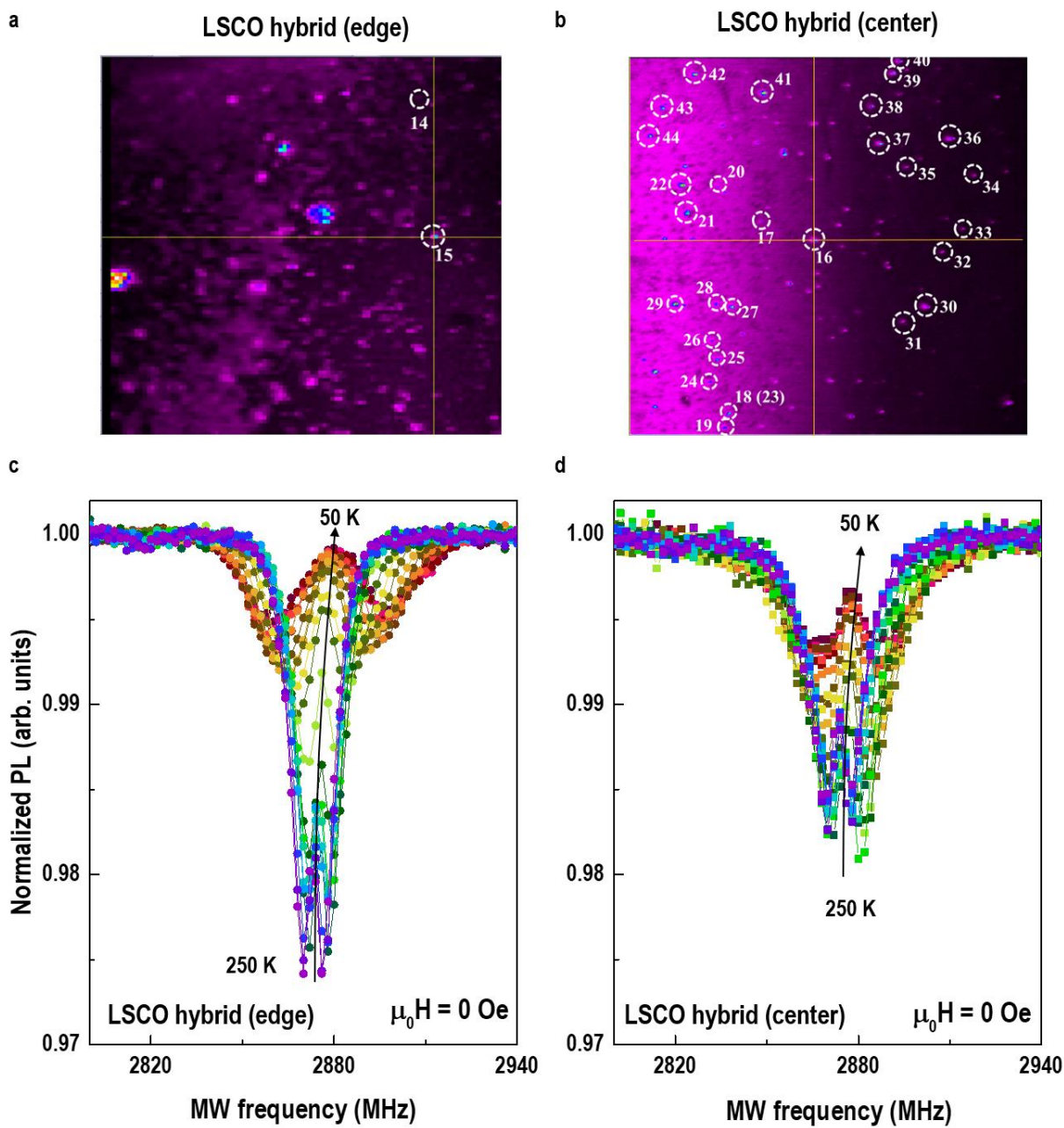

**Fig. S18. Diamond NV magnetometry on LSCO hybrid structures**. Nanodiamonds (NDs) with ensemble NV centers were dispersed randomly on the surface of LSCO hybrid structures on LAO substrates. (a) and (b) Optical microscopy images of NDs at the edge and center regions of LSCO hybrid structures. Assisted from motorized precision translation stages at the microscale, we could select specific NDs at the edge and center regions of LSCO hybrid structures and investigate their temperature dependencies. (c) and (d) Zero-field optically detected magnetic resonance (ODMR) spectra of selected NDs at the edge and center regions of LSCO hybrid structures when $T$ increases from 50 to 250 K. As increasing temperature, the splitting ($2\gamma B$) of ODMR spectra reduces. The temperature dependences of $2\gamma B$ at edge and center regions of LSCO hybrid structures are plotted in Fig. 4I of main text.

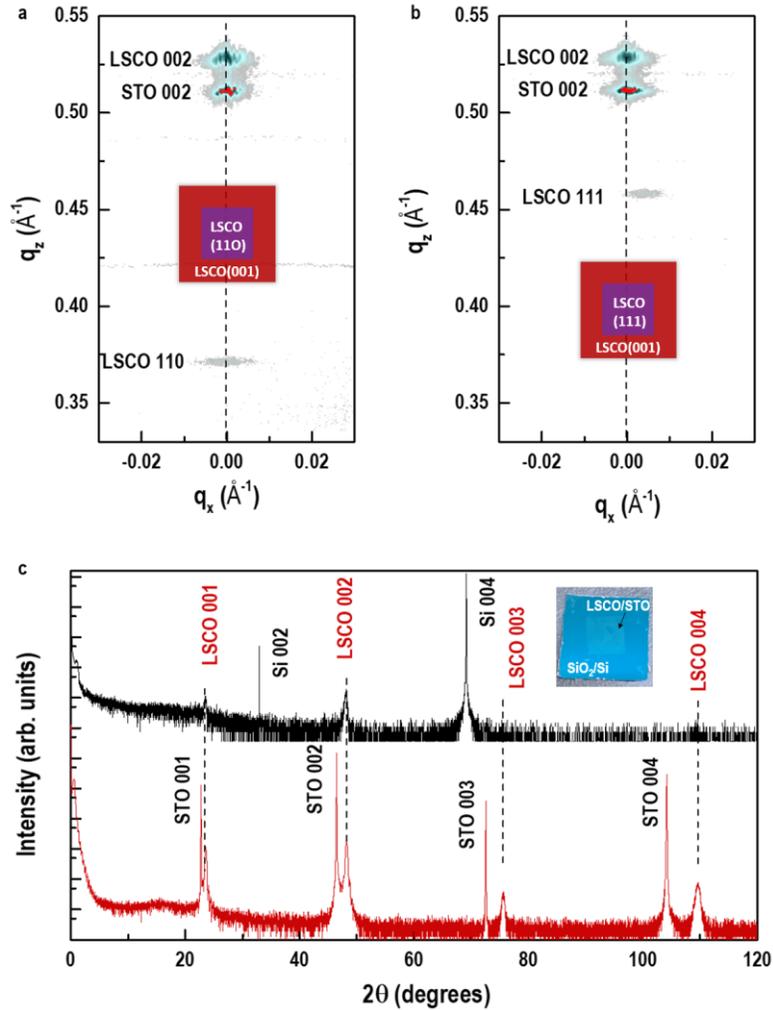

**Fig. S19. Fabrication of LSCO hybrid structures on arbitrary substrates**. We followed the same protocol as the (001)-LSCO hybrid structures grown on other single crystalline substrates. Firstly, we fabricated the (110)- and (111)-oriented STO/SAO bilayers on the relevant substrates. Then, we dissolved the SAO and support the ultrathin STO membranes using thermally-release tapes. The (110)- and (111)-oriented FS-STO membranes were transferred to the (001)-oriented STO substrates. Finally, we deposited the LSCO layers on the modified (001)-oriented STO substrates using PLD technique. (a) and (b) RSMs around 002 reflections of (110)- and (111)-oriented 30-nm-thick LSCO hybrid structures grown on the (001)-oriented STO substrates, respectively. The (110)- and (111)-reflections from FS-LSCO/STO membranes are observed except for (002) reflections of LSCO layers and STO substrates, suggesting that different oriented LSCO layers can be successfully fabricated on the (001)-oriented STO substrates. Another example is the fabrication of epitaxial (001)-LSCO layers on the SiO$_2$/Si substrates. Typically, it is challenging to grow a functional oxide film directly on the silicon or silicon oxide substrates due to the enormous misfit strain. Using a freestanding STO membrane, we were able to transfer it onto the SiO$_2$/Si substrates and then fabricated high-quality single-crystalline functional oxides. (c) XRD 2θ-ω scan of a LSCO hybrid structure on SiO$_2$/Si substrates. A XRD 2θ-ω curve of A (001)-oriented LSCO single film is plotted as a direct comparison. Only (00*l*) reflections of LSCO layers are observed, suggesting the epitaxial growth of LSCO layers. These results indicate that the method in present work is readily applicable to any arbitrary substrates.